\documentclass[twocolumn]{aastex63}

\usepackage{graphicx}	
\usepackage{color}
\usepackage{amsmath}	% Advanced maths commands
\usepackage{amssymb}	% Extra maths symbols
\usepackage{mathrsfs}

\usepackage{bigints}

\newcommand{\msun}{M_\odot}
\newcommand{\msunyr}{M_\odot~{\rm yr}^{-1}}
\newcommand{\zsun}{Z_\odot}
\newcommand{\K}{{\rm K}}

\shorttitle{GWB from high-z BBHs}
\shortauthors{K. Inayoshi, K. Kashiyama, E. Visbal \& Z. Haiman}

\begin{document}

\title{Gravitational wave backgrounds from coalescing black hole binaries at cosmic dawn: \\an upper bound
}

\correspondingauthor{Kohei Inayoshi}
\email{inayoshi@pku.edu.cn}

\author{Kohei Inayoshi}
\affiliation{Kavli Institute for Astronomy and Astrophysics, Peking University, Beijing 100871, China}

\author{Kazumi Kashiyama}
\affiliation{Research Center for the Early Universe, Graduate School of Science, University of Tokyo, Bunkyo-ku, Tokyo 113-0033, Japan}
\affiliation{Department of Physics, Graduate School of Science, University of Tokyo, Bunkyo-ku, Tokyo 113-0033, Japan}

\author{Eli Visbal}
\affiliation{
Department of Physics and Astronomy and Ritter Astrophysical Research Center,
University of Toledo, Toledo, Ohio 43606}

\author{Zolt\'an Haiman}
\affiliation{Department of Astronomy, Columbia University, New York, NY 10027}

\begin{abstract}

The successive discoveries of binary merger events by Advanced LIGO-Virgo 
have been revealing the statistical properties of binary black hole (BBH) populations.
A stochastic gravitational wave background (GWB) is a useful tool to probe the cosmological evolution of 
those compact mergers.
In this paper, we study the upper bound on a GWB produced by BBH mergers, 
whose stellar progenitors dominate the reionization process at the cosmic dawn.
Since early reionization by those progenitors yields a high optical depth 
of the universe inconsistent with the {\it Planck} measurements,
the cumulative mass density is limited to $\rho_\star \la 10^7~\msun~{\rm Mpc}^{-3}$.
Even with this upper bound,
the amplitude of a GWB owing to the high-$z$ BBH mergers is expected to be as high as
$\Omega_{\rm gw}\simeq 1.48_{-1.27}^{+1.80}\times 10^{-9}$ at $f\simeq 25$ Hz,
while their merger rate at the present-day is consistent or lower than the observed GW event rate.
This level of GWB is detectable at the design sensitivity of Advanced LIGO-Virgo and would indicate
a major contribution of the high-$z$ BBH population to the local GW events.
The spectral index is expected to be substantially flatter than the canonical value of $\simeq 2/3$
generically produced by lower-redshift and less massive BBHs.
Moreover, if their mass function is more top-heavy than in the local universe,
the GWB spectrum is even more skewed toward lower frequencies,
which would allow us to extract information on the mass function of 
merging BBHs at high redshifts.

\end{abstract}

\keywords{Gravitational waves -- Reionization -- Compact objects -- Cosmic background radiation\vspace{5mm}}

%%%%%%
% Sec. 1 %
%%%%%%

\section{Introduction}
\label{sec:intro}

Since the detections of gravitational waves (GWs) associated with compact binary mergers
have opened a new window to explore our universe, the number of GW sources has been increasing
substantially \citep{LIGO_back_2016,Abbott_2016_Astro,Abbott_PRL_2016}.
Recently, a new catalog of 47 compact binary mergers including 44 binary black holes (BBHs) detected 
in Advanced LIGO-Virgo observing runs O3a has been reported \citep{LIGO_GWTC_2020}.
With the updated sample, the estimation of the primary BH mass spectrum and BBH merger rate have
been substantially improved.

The origin of such massive BBHs and their formation pathway have been extensively discussed
based on the properties of detected BBHs (e.g., the distribution of mass and spin components).
So far, various models have been proposed; through massive binary evolution in low-metallicity environments 
\citep{Dominik_2012,K14,K16,Belczynski_2016_a,Inayoshi_2017,vandenHeuvel_2017,Neijssel_2019,Santoliquido_2021}, 
dynamical processes in dense stellar clusters and galactic nuclei
\citep{PortegiesZwart_2000,Rodriguez_2015,OLeary_2016,Stone_2017,Mapelli_2016,Bartos_2017,McKernan_2018,
Tagawa_2020}, and primordial BH formation \citep{Nakamura_1997,Sasaki_2016, Ali-Haimoud_2017}.

Referring to the redshift-dependent BBH merger rate, a larger number of BBHs would merge at earlier epochs
and thus most of the individually unresolved mergers produce a GW background (GWB)
\citep{LIGO_back_2016,Kowalska_2015,Inayoshi_2016,Hartwig_2016, Dvorkin_2016, Callister_2020, 
Perigois_2020,LIGO_GWB_2021}.
The detection of a GWB will be used to probe the formation epoch and efficiency of coalescing BBHs,
constrain the mass function for massive star/BH populations initiated in the early universe,
and even provide information on the history of cosmic reionization.
More specifically, the existence of high-$z$, massive BBH populations (e.g., the remnant BHs of 
Population III stars; hereafter PopIII stars) expected to typically form at $z\sim 10-30$
would produce a GWB detectable by LIGO/Virgo 
with a unique spectral shape that flattens significantly at $\sim 30$ Hz, 
which is distinguishable from the spectral index of $\sim 2/3$ generically produced by lower redshift 
and less-massive BBHs \citep{Inayoshi_2016}.
A recent population synthesis study also claimed a deviation of the spectral index from the 
canonical value if the PopIII contribution is included \citep{Perigois_2020}.

Massive stellar progenitors of merging BBHs formed at the cosmic dawn ($z\ga 6$) are also efficient producers 
of ionizing radiation in the early universe and are expected to dominate the reionization process. 
Recently, {\it Planck} has reported an updated estimate of the optical depth of the universe to 
electron scattering inferred from the cosmic microwave background (CMB) anisotropies;
$\tau_{\rm e}\simeq 0.052\pm 0.008$ \citep{Planck_2020}. 
This low value would give a stringent constraint on the star formation history and the total stellar mass 
budget available for BBH formation at higher redshifts \citep{Visbal_2015,Inayoshi_2016}.
Therefore, this constrains the amplitude of a GWB owing to BBH mergers originating from high-$z$ populations.

In this paper, we study the upper bound of the GWB produced by BBH mergers 
taking into account the constraint on the cumulative stellar mass from cosmic reionization.
We find that even with the upper bound, the GWB signal is still detectable at the Advanced LIGO-Virgo design sensitivity,
while the merger rate at $z\simeq 0$ is consistent or lower than the observed GW event rate.
Using the updated BBH properties from the LIGO-Virgo O3a observing run and the new value of $\tau_{\rm e}$, 
we infer a GWB spectral shape with a characteristic flattening, which is even more skewed toward lower frequencies 
if the mass function is more top-heavy than in the local universe.
This is also an updated study on our previous paper \citep{Inayoshi_2016} published after the detection 
of the first source GW150914, in which a single value of the BH mass was assumed and
the higher optical depth ($\tau_{\rm e}\simeq 0.06\pm 0.016$) provided by the previous Planck estimate \citep{Planck_2015}.

The rest of this paper is organized as follows: in \S\ref{sec:upper}, we describe our reionization model and provide 
an upper bound on the stellar mass density at the cosmic dawn, consistent with the recent Planck result.
In \S\ref{sec:merger}, we calculate the redshift-dependent merger rate of BBHs
under the constraint from the reionization history.
In \S\ref{sec:GWB}, we present the expected GWB spectra for various BH mass distributions, 
and discuss the detectability of those GWB signals and possible implications.
Finally, in \S\ref{sec:summary}, we summarize the conclusion of this paper.
Throughout this paper, we assume a $\Lambda$ cold dark matter cosmology consistent with the latest constraints from
Planck \citep{Planck_2020}; $h=0.6732$, $\Omega_{\rm m}=0.3158$, $\Omega_{\rm b}=0.02238$, and $Y_{\rm He}=0.247$.

\vspace{5mm}
%%%%%%
% Sec. 2 %
%%%%%%
\section{The upper bound of the stellar mass density in the cosmic dawn}
\label{sec:upper}

In this paper, we consider two BBH populations originating from different cosmic star formation histories,
which are referred to as low-$z$ and high-$z$ BBH populations, respectively.
The low-$z$ BBH population follows the ``observed" cosmic star-formation rate density (SFRD), 
which is characterized by $\dot{\rho}_\star \propto (1+z)^{2.7}$ at $z\la 2$, 
has a peak at the cosmic noon around $z\simeq 2$, and declines $\dot{\rho}_\star \propto (1+z)^{-2.9}$ 
toward higher redshifts \citep[][]{Madau_Dickinson_2014}.
This SFRD is often used for estimating the merger rates of compact binaries in many previous studies in literature
\citep[e.g.,][]{LIGO_back_2016,LIGO_GWB_2021}.

The observed SFRD of the low-$z$ stellar population is not sufficient to reionize the universe by $z\sim 6$ 
and to then keep it ionized \citep[e.g.,][]{Robertson_2015}\footnote{\cite{Robertson_2015} computed the SFRD by extrapolating 
the actual observed luminosity function to a faint, unobserved value of $L_{\rm min} = 0.001~L_\star$, where $L_\star$ is the characteristic 
luminosity of each parameterization, e.g., Schechter or broken power-law models
(see also discussion by \citealt{Madau_Dickinson_2014}).},
unless a large fraction ($\ga 20\%$) of ionizing photons
can escape from galaxies to the intergalactic media \citep{Madau_Dickinson_2014} or the production efficiency 
of ionizing photons is sufficiently high (e.g., massive stars with stripped envelopes via binary interactions; see \citealt{Ma_2016}).
Therefore, a stellar population formed in fainter, undetected galaxies must exist beyond $z>6$;
the star formation rate extends to higher redshifts and is responsible for the completion of cosmic reionization by $z=z_{\rm reion}$.
In this paper, the high-$z$ BBH population refers to BBHs originating from such high-$z$ stellar components.
Their star formation activity is expected to take place in metal-poor/low-metallicity environments, 
e.g., protogalaxies in dark-matter (DM) halos with virial temperatures of $T_{\rm vir} \simeq 10^3-10^4~\K$
\citep{Bromm_Yoshida_2011, Wise_2012}.
Although the cosmic SFRD at $z\ga z_{\rm reion}$ has not been constrained tightly by direct observations of 
star-forming high-$z$ galaxies, the measurements of the optical depth of the universe to electron scattering 
imprinted into the CMB anisotropies would give a constraint on the total stellar
mass budget available for BBH formation at higher redshifts \citep{Visbal_2015,Inayoshi_2016}.
In this section, we give the upper bound of the total (comoving) mass density of stars at $z\ga z_{\rm reion}$,
depending on the physical parameters related to reionization processes.

\vspace{2mm}
\subsection{The semi-analytical cosmic reionization model}
\label{sec:reionmodel}

We describe the redshift-dependent cosmic SFRD at $z\geq z_{\rm reion}$
using a phenomenological model with three fitting parameters:
\begin{equation}
\dot{\rho}_\star(z) =\frac{a_p}{1+[(1+z)/b_p]^{c_p}}~~~~~{\rm at} ~z\geq z_{\rm reion}.
\label{eq:app1}
\end{equation}
This parameterization is motivated by the functional form used in \cite{Madau_Dickinson_2014},
except the decline toward lower redshifts.
If ionizing photons from star-forming galaxies lead the reionization process,
the photon production rate per comoving volume is proportional to $\dot{\rho}_\star (z)$ as
\begin{equation}
\dot{n}_{\rm ion} = \frac{f_{\rm esc} \eta_{\rm ion} \dot{\rho}_\star(z)}{m_{\rm p}},
\end{equation}
where $f_{\rm esc}$ is the escape fraction of ionizing photons from galaxies to the intergalactic media (IGM),
$\eta_{\rm ion}$ is the ionizing photon number per stellar baryon, and $m_{\rm p}$ is the proton mass.
Evidently, the two quantities have different values for each population depending on the typical properties 
of their host DM halos, initial mass function, and metallicity \citep[see][and references therein]{Yung_III_2020,Yung_IV_2020}.
Following previous studies \citep[e.g.,][]{Greif_Bromm_2006,Johnson_2013,Visbal_2020,Liu_Bromm_2020},
we adopt fiducial values of $f_{\rm esc} = 0.1$ and $\eta_{\rm ion}=4\times 10^3$ 
for stellar populations that form in early protogalaxies and dominate the reionization process
(e.g., \citealt{Wise_2014}; $f_{\rm esc}\simeq 0.1$ for DM halos with $\ga 10^8~\msun$).
Note that the value of $\eta_{\rm ion}$ is consistent with that of a $Z\simeq 0.02~\zsun$ stellar population 
(hereafter, PopII)\footnote{
Relatively metal-enriched stellar populations with smaller $\eta_{\rm ion}$ could contribute to the reionization process
at $z>6$ if the metallicity of galaxies decline toward high-redshifts as weakly as seen at $z\simeq 2-3$ \citep{Sanders_2020,Suzuki_2021}.}
which follows a Salpeter IMF with a mass range of $0.1-100~\msun$.
Since these values are uncertain, we also discuss the dependence on the product 
$f_{\rm esc}\eta_{\rm ion}$ that matters rather than their individual values\footnote{
Recent sub-millimeter observations 
have revealed the existence of star-forming massive galaxies at $z>2$ that are highly obscured by dust grains
\citep[e.g.,][]{Wang_2019,Gruppioni_2020}.
This galaxy population might not significantly contribute to the reionization process (presumably smaller values of $f_{\rm esc}$), 
but they still would produce a large amount of stars that potentially constitute another source of BBHs \citep{Boco_2019,Boco_2021}.
Throughout this paper, however, their contribution to the BBH merger rate is not explicitly considered.}.

With the photon production rate, the IGM ionized volume fraction $Q_{\rm H_{II}}(z)$ is calculated 
by the differential equation \citep[e.g.,][]{Haiman_Loeb_1997,Madau_1999,Wyithe_Loeb_2003,Haiman_Bryan_2006};
\begin{equation}
\frac{dQ_{\rm H_{II}}}{dt} = \frac{\dot{n}_{\rm ion}}{\langle n_{\rm H}\rangle}
-\frac{Q_{\rm H_{II}}}{t_{\rm rec}},
\end{equation}
where the IGM recombination time is given by
\begin{equation}
t_{\rm rec} = \left[C_{\rm H_{II}} \alpha_{\rm B}\left(1+\frac{Y_{\rm He}}{4X_{\rm H}}\right)
\langle n_{\rm H}\rangle(1+z)^3\right]^{-1},
\end{equation}
$\alpha_{\rm B}$ is the case B recombination coefficient at an IGM temperature of 
$T = 2\times 10^4~\K$, $\langle n_{\rm H}\rangle$ is the IGM mean comoving number density of hydrogen, 
$C_{\rm H_{II}}\equiv \langle n_{\rm HII}^2 \rangle / \langle n_{\rm HII} \rangle ^2 $ is a clumping factor of ionized hydrogen,
and $X_{\rm H}=0.76$ and $Y_{\rm He} = 0.24$ 
are the hydrogen and helium mass fractions.
We adopt $C_{\rm H_{II}}=4$ \citep[e.g.,][]{Pawlik_2009, Robertson_2015}.
Finally, for a given reionization history associated with an SFRD model, the optical depth $\tau_{\rm e}$ 
is calculated with
\begin{equation}
\tau_{\rm e}(z) = c\langle n_{\rm H}\rangle \sigma_{\rm T}\int_0^z
Q_{\rm H_{II}}(z')\frac{(1+z')^2}{H(z')} \left(1+\frac{\eta_{\rm He}Y_{\rm He}}{4X_{\rm H}}\right)
dz',
\end{equation}
where $c$ is the speed of light, $\sigma_{\rm T}$ is the cross section of Thomson scattering, and $H(z)$ is the Hubble parameter.
Helium is assumed to be singly ionized with hydrogen at $z \geq 3$ ($\eta_{\rm He}=1$), but be doubly ionized at the lower redshifts
($\eta_{\rm He}=2$).

%%%%%%%
%    Fig. 1   %
%%%%%%%
\begin{figure*}
 \begin{minipage}{0.5\hsize}
  \begin{center}
   \includegraphics[width=85mm]{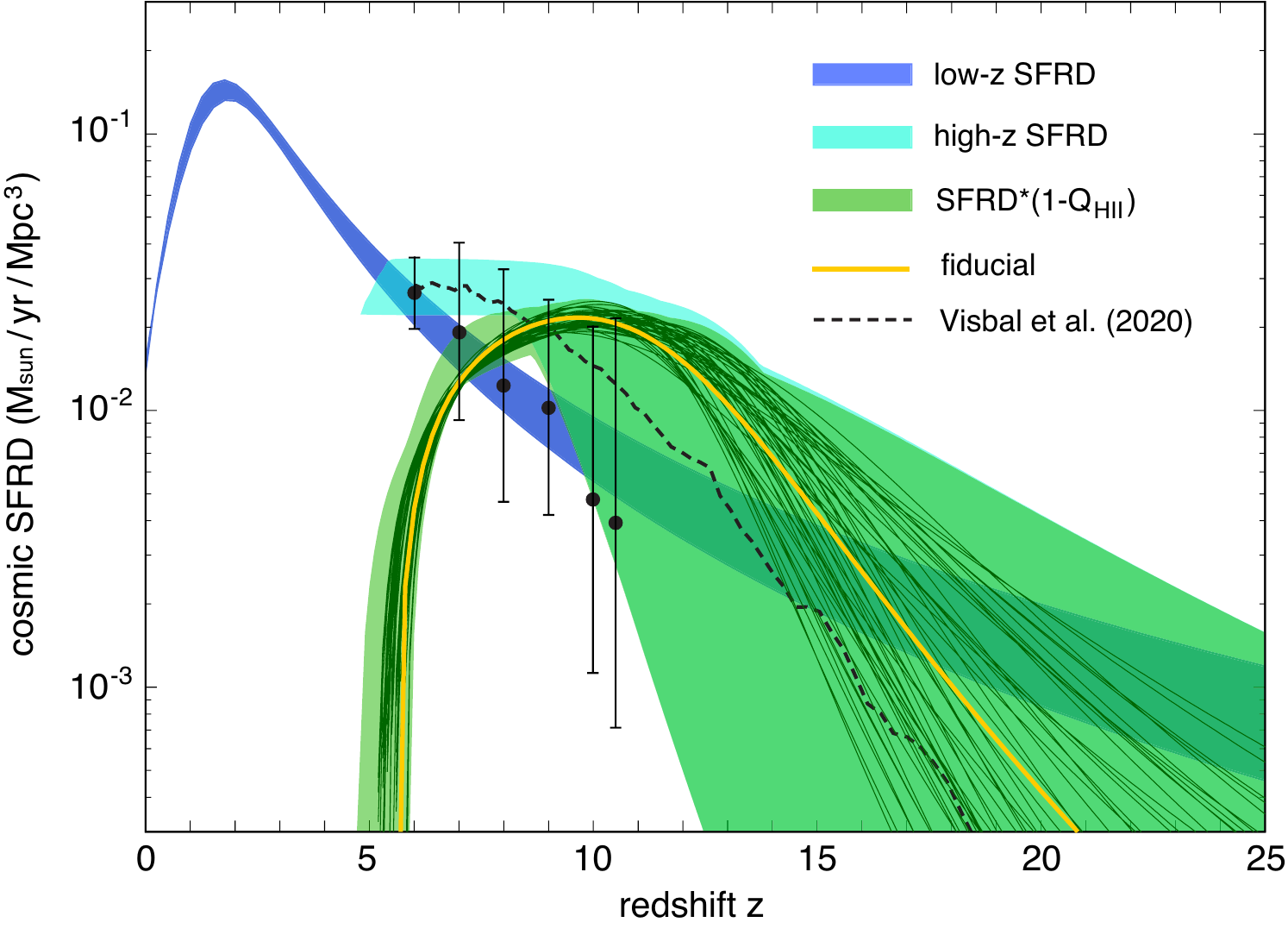}
  \end{center}
   \end{minipage}
   \hspace{-2mm}
 \begin{minipage}{0.5\hsize}
  \begin{center}
   \includegraphics[width=86mm]{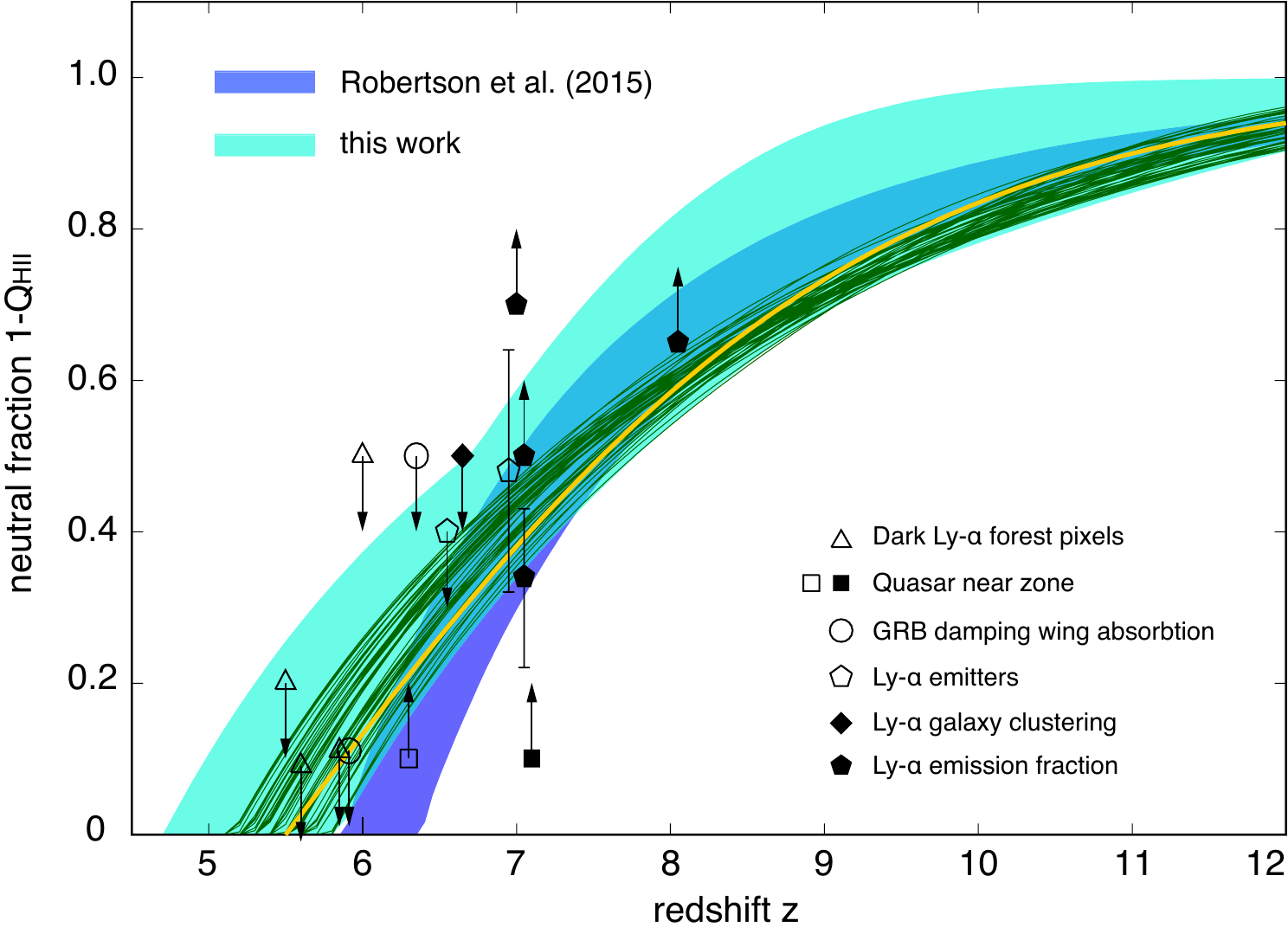}
  \end{center}
  \end{minipage}
  \vspace{-2mm}
\caption{{\it Left panel}: cosmic SFRDs of the stellar population at $z\ga 5$ (light blue region), for which the Planck result
($\tau_{\rm e} = 0.0522 \pm 0.0080$ and $z_{\rm reion}^{50\%} = 7.50 \pm 0.82$) are consistently reproduced.
The green region presents the SFRD in neutral regions before the completion of reionization,
and the solid curves show 50 cases that form stars with $\rho_\star \simeq 10^7~\msun~{\rm Mpc}^{-3}$.
Our fiducial SFRD model is highlighted with the yellow curve.
For references, we overlay other SFRDs for the high-$z$ population \citep[black dashed;][]{Visbal_2020} and 
for the low-$z$ population (blue region; \citealt{Madau_Dickinson_2014}, \citealt{Robertson_2015})
together with the errors of the observed SFRD at $z\ga 6$ \citep{Robertson_2015}.
{\it Right panel}: evolution of cosmic volume fraction of neutral regions $1-Q_{\rm H_{II}}(z)$ in our model 
(light blue region, green curves, and yellow curve),
together with the observational constraints compiled by \cite{Robertson_2015} and the result consistent with 
the previous Planck estimate (blue region).
}
\label{fig:app1}
\vspace{4mm}
\end{figure*}

In this framework, we investigate the ranges of the three parameters ($a_p$, $b_p$, and $c_p$) in Eq.~(\ref{eq:app1}),
which are constrained from (i) the SFRD estimated from UV luminosities at $z\simeq 6$
\citep{Robertson_2015}, (ii) the Planck measured optical depth $\tau_{\rm e} = 0.0522 \pm 0.0080$, 
and (iii) reionization redshift mid-point $z_{\rm reion}^{50\%} = 7.50 \pm 0.82$ \citep{Planck_2020}. 
%Throughout the argument below, the most updated cosmological parameters are adopted;
%$h=0.6732$, $\Omega_{\rm m}=0.3158$, $\Omega_{\rm b}=0.02238$, and $Y_{\rm He}=0.247$
%\citep{Planck_2020}.
We note that those values of $\tau_{\rm e}$ and $z_{\rm reion}^{50\%}$ are estimated 
for a specific shape of $Q_{\rm H_{II}}(z)$ and the resultant shape obtained from our semi-analytical model 
is similar to those assumed by the Planck team.
A recent paper by \cite{Ahn_Shapiro_2020} showed with a suite of models of early reionization due to PopIII stars
(see also our discussion in \S\ref{sec:pop3}), their best fit models to the Planck polarization data go up to $\tau_{\rm e} \simeq 0.064$.
Therefore, the optical depth we adopt is a conservative choice.

\vspace{2mm}
\subsection{The upper bound of the total stellar mass}
\label{sec:rhoII}

First, we consider the case where a single stellar population dominates the reionization process,
that is, a single value of $f_{\rm esc}\eta_{\rm ion}(=400)$ is adopted at all redshifts.
In the left panel of Fig.~\ref{fig:app1}, we show the range of cosmic SFRDs at $z\ga 5$, for which the Planck measured values of 
$\tau_{\rm e}$ and $z_{\rm reion}^{50\%}$ are consistently reproduced (light-blue region).
Those SFRDs are as high as $\dot{\rho}_\star \sim (2-4)\times 10^{-2}~\msunyr~{\rm Mpc}^{-3}$ at $z\la 9$
and begin to decline toward high redshifts at $z\ga 10-14$.
All the cases shown here are consistent with the observed SFRDs within the errors over $5\la z\la 10.5$
and smoothly connect to the SFRD measured at lower redshifts (blue region; 
\citealt{Madau_Dickinson_2014} and \citealt{Robertson_2015}).
For comparison, we overlay an SFRD model calculated by \cite{Visbal_2020}, where more realistic prescriptions
for star formation, radiation feedback, IGM metal pollution, and the transition from PopIII to PopII stars are considered.
In the right panel of Fig.~\ref{fig:app1}, we present the neutral fraction $(1-Q_{\rm H_{II}})$ of the IGM as a function of redshift,
together with the observational constraints compiled by \cite{Robertson_2015} (and references therein).
The computed reionization history is overall consistent with these observational results.
The result consistent with the previous Planck estimate (blue region) is overlaid for comparison.

%%%%%%%
%    Fig. 2   %
%%%%%%%
\begin{figure*}
 \begin{minipage}{0.5\hsize}
  \begin{center}
   \includegraphics[width=83mm]{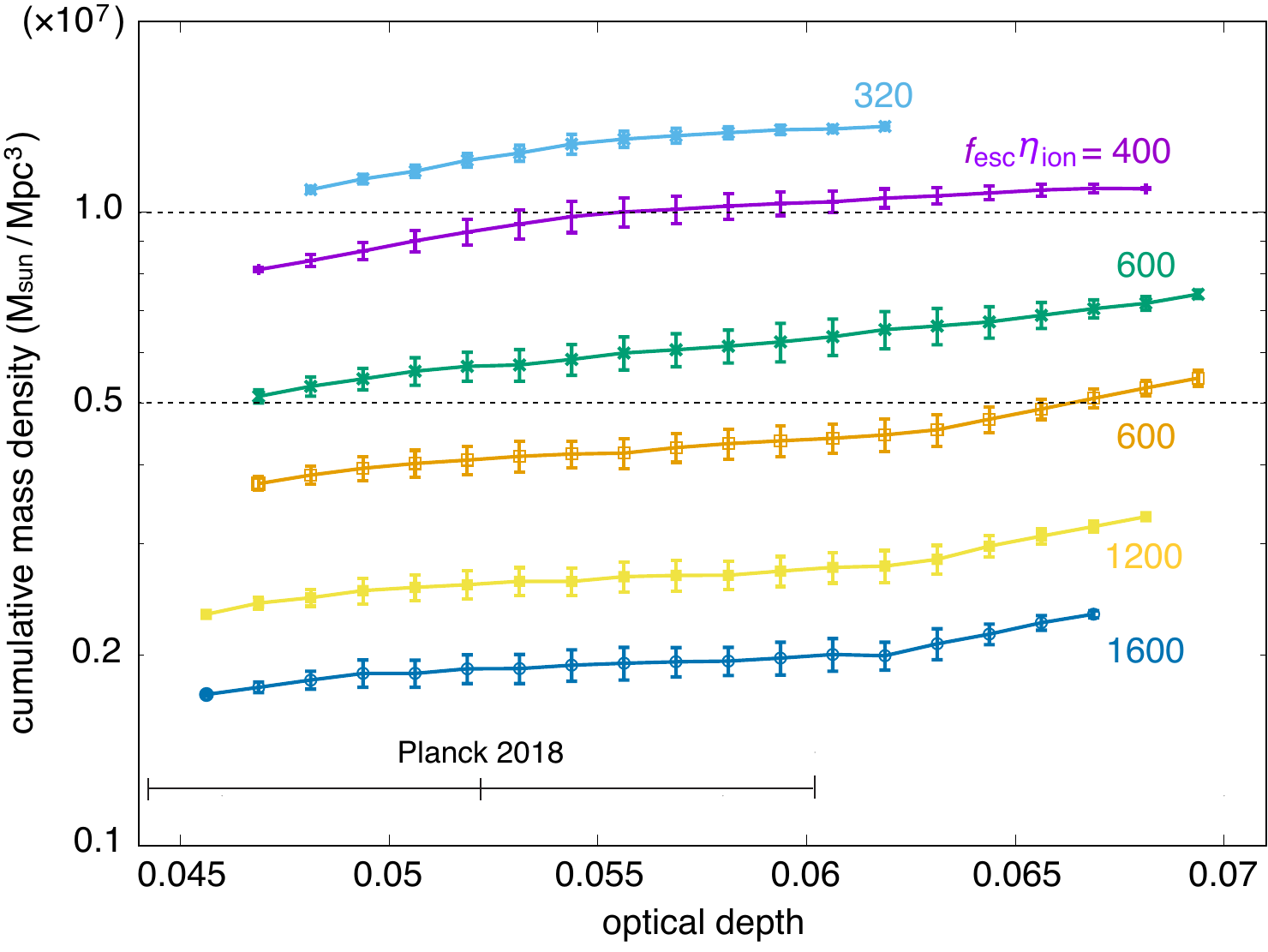}
  \end{center}
   \end{minipage}
 \begin{minipage}{0.5\hsize}
  \begin{center}
   \includegraphics[width=83mm]{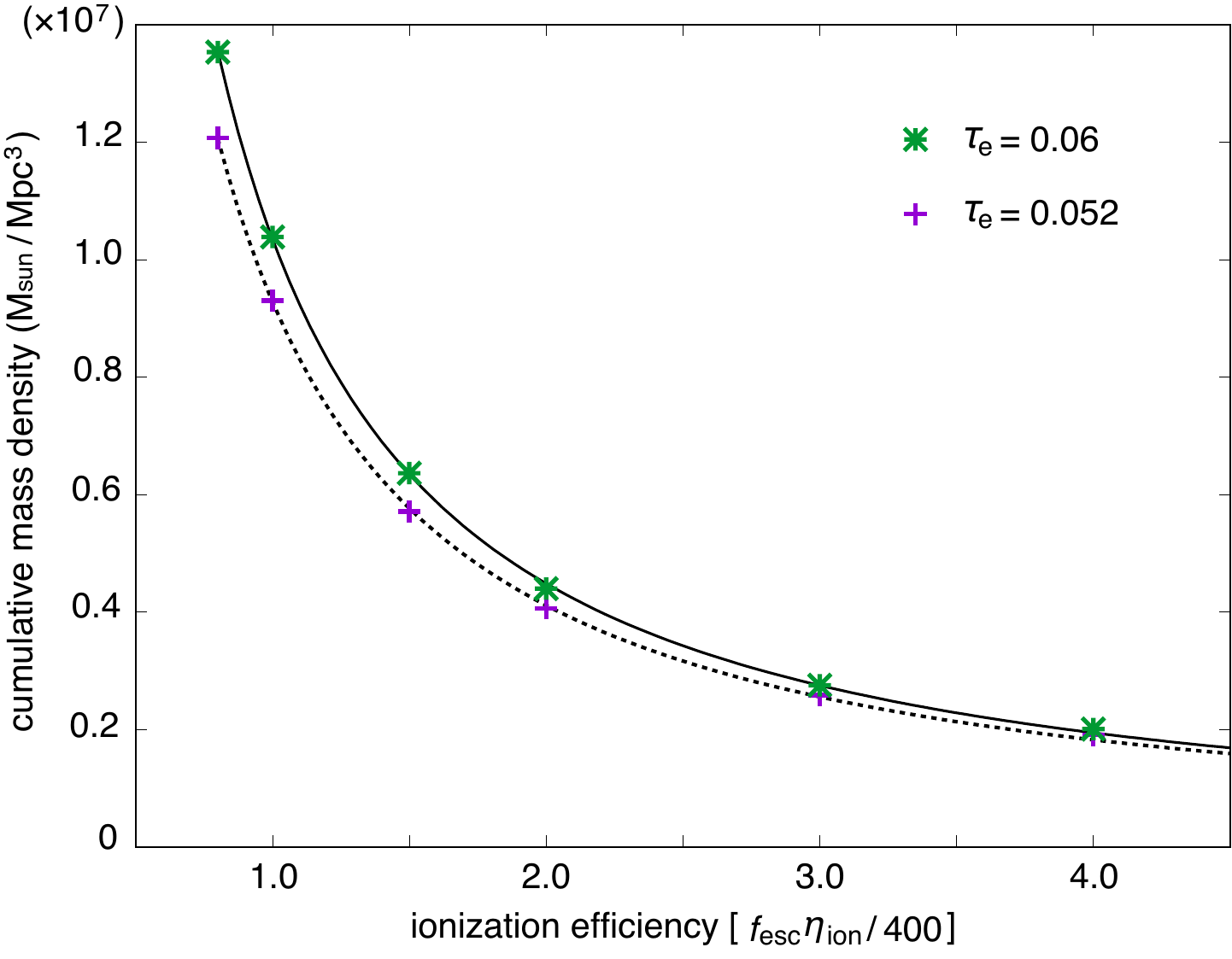}
  \end{center}
  \end{minipage}
  \vspace{-2mm}
\caption{{\it Left panel}: cumulative stellar mass density consistent with the Planck measurements
as a function of the ionization efficiency; $f_{\rm esc}\eta_{\rm ion} = 320$,
$400$ (fiducial), $600$, $800$, $1200$, and $1600$ from the top to the bottom
(the $1\sigma$ errors are shown in each $\tau_{\rm e}$ bin with a size of $1.25\times 10^{-3}$).
The range of $\tau_{\rm e}$ inferred from the Planck observation is shown \citep{Planck_2020}.
{\it Right panel}: the relation between $\rho_\star$ and $ f_{\rm esc}\eta_{\rm ion}$ is shown 
for the two cases of $\tau_{\rm e}=0.052$ and $0.06$. The results are well approximated by Eq.~(\ref{eq:rhostar_app}).
}
  \label{fig:app2}
  \vspace{2mm}
 \end{figure*}

In this paper, we consider a GWB produced by stellar populations that contribute to cosmic reionization
and presumably form in DM halos with $T_{\rm vir} \la 2\times 10^4~\K$, 
where gas is vulnerable to photoionization heating feedback \citep[e.g.,][]{Dijkstra_2004,Okamoto_2008}.
Therefore, we assume that the formation of the early component is suppressed in ionized regions and thus
takes place only in neutral regions with a cosmic volume fraction of $1-Q_{\rm H_{II}}(z)$\footnote{
There would exist metal-free DM halos that are massive enough to overcome the photoionization heating feedback 
and make PopIII stars even in ionized regions of the IGM after reionization \citep{Johnson_2010,Kulkarni_2019}.
Although such a formation pathway of PopIII stars is allowed without violating the Planck constraint,
we here do not consider their remnant (binary) BHs as the high-$z$ population.}.
The SFRDs of such populations are shown with the green shaded region, which peaks around 
$z\simeq z_{\rm reion}^{50\%}$ and sharply drops at $z\simeq z_{\rm reion}$.
For each model, we calculate the cumulative stellar mass density defined by
\begin{equation}
\rho_\star = \int_{z_{\rm reion}}^\infty \dot{\rho}_\star(z)\{1-Q_{\rm H_{II}}(z)\} \frac{dt}{dz}dz.
\end{equation}
In the left panel of Fig.~\ref{fig:app2}, we present the cumulative stellar mass density as a function 
of $\tau_{\rm e}$ for the cases with different values of
$\eta_{\rm ion} f_{\rm esc} =320$, $400$ (fiducial), $600$, $800$, $1200$, and $1600$ from the top to the bottom
(the $1\sigma$ errors are shown in each $\tau_{\rm e}$ bin with a size of $1.25\times 10^{-3}$).
Note that $z_{\rm reion}^{50\%} = 7.50 \pm 0.82$ is satisfied for all the cases.
For the fiducial case, the cumulative mass density is as high as $\rho_\star \simeq 10^7~\msun~{\rm Mpc}^{-3}$
and depends on the optical depth as $\propto \tau_{\rm e}^{0.68}$.
Within the uncertainty of $\tau_{\rm e}=0.0522\pm0.008$ \citep{Planck_2020}, 
the value of $\rho_\star$ varies within $\simeq 0.1$ dex.
With higher values of $f_{\rm esc} \eta_{\rm ion}$, the cumulative mass density decreases 
so that the resultant optical depth becomes consistent with that measured by the Planck.
In the right panel of Fig.~\ref{fig:app2}, we show the dependence of $\rho_\star$ on $f_{\rm esc} \eta_{\rm ion}$ 
for the two cases of $\tau_{\rm e}=0.052$ and $0.06$.
For both cases, the results are well fitted with a single power law of 
$\rho_\star/(10^7~\msun~{\rm Mpc}^{-3}) = a_q(f_{\rm esc}\eta_{\rm ion} /400)^{-b_q}$,
where $a_q=1.04~(0.929)$ and $b_q=1.21~(1.18)$ for $\tau_{\rm e}=0.06~(0.052)$.
Therefore, we obtain the relation between the total stellar mass density formed 
by the end of reionization and 
the physical parameters of the reionization process
\begin{align}
\rho_\star & \simeq 1.04\times 10^7~\msun~{\rm Mpc}^{-3}\nonumber\\
&~~~\times 
\left(\frac{f_{\rm esc}}{0.1}\right)^{-1.2}
\left(\frac{\eta_{\rm ion}}{4\times 10^3}\right)^{-1.2}
\left(\frac{\tau_{\rm e}}{0.06}\right)^{0.68}.
\label{eq:rhostar_app}
\end{align}
We note that the mass density is broadly consistent with cosmological hydrodynamical simulations 
for high-$z$ galaxy formation \citep[e.g.,][]{Johnson_2013}.
The value in Eq.~(\ref{eq:rhostar_app}) is considered to be the {\it upper bound} of the stellar mass
formed at $z\ga z_{\rm reion}$
since it would be lowered if other rarer but more intense radiation sources (e.g., 
metal-free PopIII stars and high-$z$ quasars) could contribute to reionization
\citep[][see also \S\ref{sec:pop3}]{Visbal_2015,Dayal_2020}.

Among all the SFRD models consistent with the Planck result,
we show 50 cases with $\rho_\star = 1.0^{+0.1}_{-0.1}\times 10^7~\msun~{\rm Mpc}^{-3}$
(green thin curves in Fig.~\ref{fig:app1}), which are characterized with a functional form of 
\begin{align}
\dot{\rho}_{\star,\rm reion}(z) = \dfrac{a_p\cdot \tanh[(z-z_{\rm reion})/d_p]}{1+[(1+z)/b_p]^{c_p}}
\label{eq:fid_SFRD}
\end{align}
at $z\geq z_{\rm reion}$, where we fit the evolution of $1-Q_{\rm H_{II}}(z)=\tanh [(z-z_{\rm reion})/d_p]$, consistent with the Planck analysis.
In this paper, we adopt one of them as our fiducial SFRD model with 
$a_p = 0.032~\msunyr~{\rm Mpc}^{-3}$, $b_p=13$, $c_q=9$, $d_p=3.74$, and $z_{\rm reion}=5.5$, 
yielding $\rho_\star = 1.0\times 10^7~\msun~{\rm Mpc}^{-3}$ and $\tau_{\rm e}=0.06$ (yellow curves in Fig.~\ref{fig:app1}).

\vspace{2mm}
\subsection{The upper bound of the PopIII stellar mass}
\label{sec:pop3}

Next, we consider the impact of metal-free PopIII stars on the reionization history and 
give the upper bound of their total mass formed across cosmic time.
PopIII stars are predicted to be more efficient at producing ionizing radiation than metal-enriched PopII stars 
\citep{Schaerer_2002,Schaerer_2003}.
If PopIII stars form with a top-heavy IMF, the ionization efficiency is substantially enhanced
(e.g., $\eta_{\rm ion}=5.1\times 10^4$ for a Salpeter IMF with $10-100~\msun$).
Moreover, a fraction of PopIII stars would form in less massive DM halos with $\la 10^7~\msun$, 
where the escape fraction of ionizing photons is expected to be as high as $f_{\rm esc}\simeq 0.5$.
Therefore, PopIII stars would affect the reionization history and create an early partial reionization, 
which leads to a higher optical depth inconsistent with the Planck result \citep{Visbal_2015}.

We repeat the same calculations but considering an effective ionization efficiency 
$\langle f_{\rm esc} \eta_{\rm ion}\rangle$ that includes the contribution of ionizing radiation
from both PopII and PopIII stars defined by
\begin{align}
\langle f_{\rm esc} \eta_{\rm ion} \rangle \dot{\rho}_\star =  
f_{\rm esc}^{\rm II} \eta_{\rm ion}^{\rm II} \dot{\rho}_{\star, \rm II}+
f_{\rm esc}^{\rm III} \eta_{\rm ion}^{\rm III} \dot{\rho}_{\star, \rm III},
\end{align}
where $\dot{\rho}_{\star, \rm III(II)}$, $f_{\rm esc}^{\rm II(III)}$, and $\eta_{\rm ion}^{\rm II(III)}$
are the cosmic SFRD, escape fraction, and number of ionizing photons per stellar baryon
for the PopII(III) population.
When the PopII population dominates the total SFRD (i.e., $\dot{\rho}_\star \simeq \dot{\rho}_{\star,\rm II}$),
the above equation is approximated as
\begin{align}
\langle f_{\rm esc} \eta_{\rm ion} \rangle \approx f_{\rm esc}^{\rm II} \eta_{\rm ion}^{\rm II}
\left( 1 + \mathcal{F}  ~\frac{\dot{\rho}_{\star, \rm III}}{\dot{\rho}_{\star, \rm II}} \right)
\end{align}
where $\mathcal{F}\equiv f_{\rm esc}^{\rm III}\eta_{\rm ion}^{\rm III}/(f_{\rm esc}^{\rm II}\eta_{\rm ion}^{\rm II})\sim O(10-100)$,
and the ratio of $\dot{\rho}_{\star, \rm III}/\dot{\rho}_{\star, \rm II}$ tends to increase with redshift but the functional
shape depends on the prescriptions for PopIII star formation.
As a reference, \cite{Visbal_2020} shows that the ratio is well approximated
by $\dot{\rho}_{\star, \rm III}/\dot{\rho}_{\star, \rm II}\approx 0.2~(f_{\rm III}/10^{-3})[(1+z)/31]^3$ at $z<30$,
where $f_{\rm III}$ is the PopIII star formation efficiency from gas clouds.
Note that the functional form of $\dot{\rho}_{\star, \rm III}/\dot{\rho}_{\star, \rm II}$ depends on the modeling of 
PopIII star formation.
We here adopt the fiducial model in \cite{Visbal_2020}; 
see also other PopIII models described in \cite{Liu_Bromm_2020},
where the SFRD seems consistent with our model with $f_{\rm III}=10^{-3}$
but tends to be higher at higher redshifts ($z>6$), leading to a higher optical depth
even with the similar amount of PopIII stars.

%%%%%%%
%    Fig. 3   %
%%%%%%%
\begin{figure}
  \begin{center}
   \includegraphics[width=83mm]{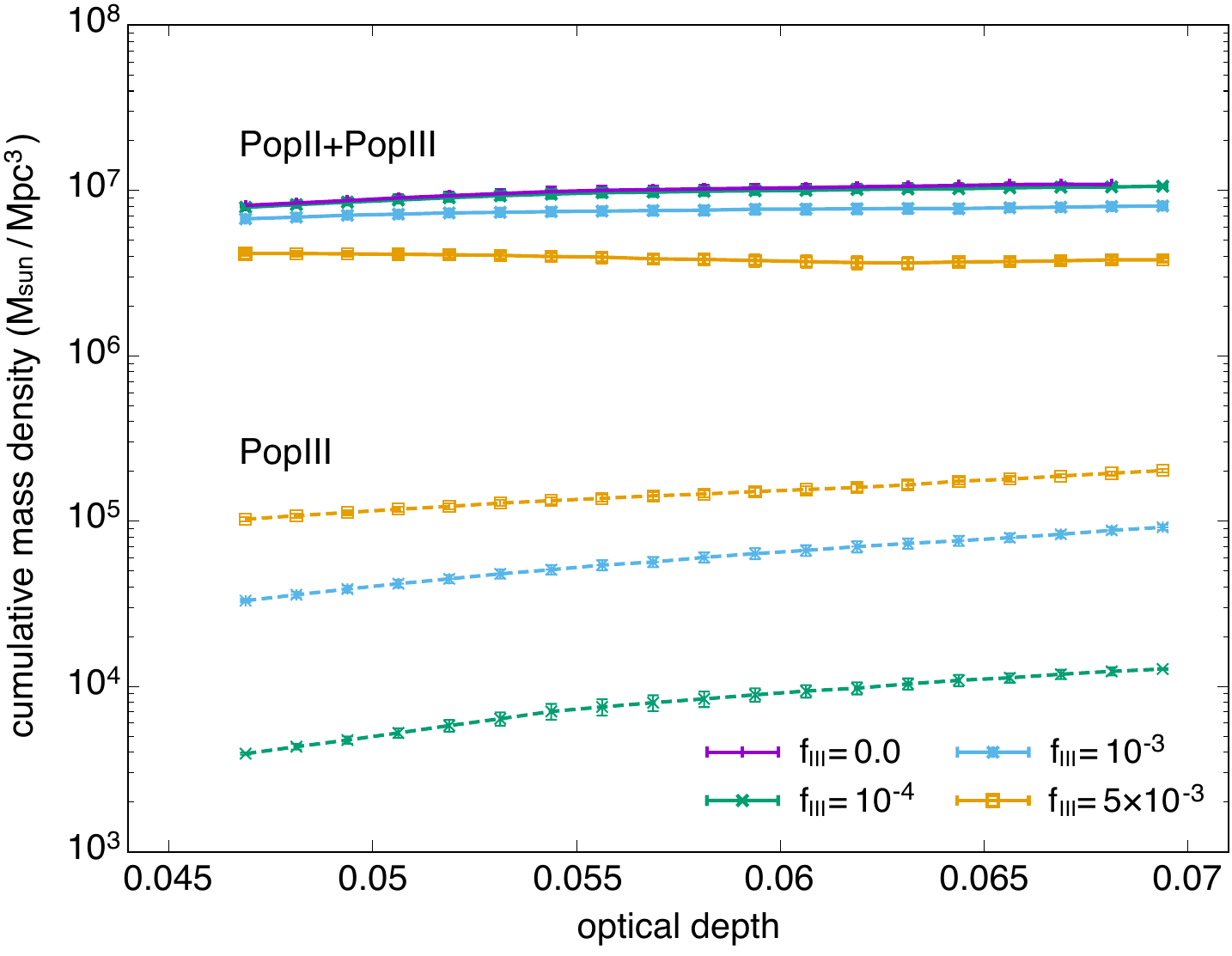}
  \end{center}
  \vspace{-4mm}
\caption{Cumulative stellar mass density of PopII+III (solid) and PopIII (dashed) stars
for different PopIII star formation efficiencies of $f_{\rm III} = 10^{-4}$, $10^{-3}$, and $5\times 10^{-3}$.
With a higher value of $f_{\rm III}$, the total amount of PopII stars gradually decreases because 
cosmic reionization is accelerated due to additional ionizing photons from PopIII stars.
In contrast, the PopIII mass density increases up to $\la 2\times 10^5~\msun~{\rm Mpc}^{-3}$.}
  \label{fig:app3}
  \vspace{1mm}
 \end{figure}

In Fig. \ref{fig:app3}, we present the cumulative stellar mass density of PopII+III (solid) and PopIII (dashed) stars
for three values of $f_{\rm III} = 10^{-4}$, $10^{-3}$, and $5\times 10^{-3}$.
We here adopt $f_{\rm esc}^{\rm II}=0.1$, $f_{\rm esc}^{\rm III}=0.5$, $\eta_{\rm ion}^{\rm II}=4\times 10^3$,
and $\eta_{\rm ion}^{\rm II}=5.1\times 10^4$.
With a higher value of $f_{\rm III}$, the total amount of PopII stars decreases gradually 
so that the total photon budget is adjusted to be consistent with the reionization history.
In contrast, the PopIII mass density increases with $f_{\rm III}$ but does not linearly scale with $f_{\rm III}$ at $\ga 10^{-3}$
because the total mass budget is regulated.
Overall, the mass density of PopIII stars is limited to $\la 2\times 10^5~\msun~{\rm Mpc}^{-3}$
and thus their contribution to the total stellar mass formed in the epoch of reionization is at most $\la 2\%$.
Note that the upper bound of the PopIII mass density is broadly consistent with the value 
estimated in \cite{Visbal_2015}, where the optical depth quoted from the Planck 2015 result 
\citep{Planck_2015} was used.

\vspace{5mm}
%%%%%%
% Sec. 3 %
%%%%%%
\section{Redshift-dependent BH merger rates}
\label{sec:merger}

\vspace{2mm}
\subsection{Properties of BBH mergers implied by LIGO/Virgo O3a run}

With the updated BBH sample in the GWTC-2 catalog \citep{LIGO_GWTC_2020}, 
the mass spectrum for the primary BH in merging binaries, $d\mathcal{R}_{\rm BBH}/dM_1$, is
found to be characterized by a broken power law with a break at $39.7^{+20.3}_{-9.1}~\msun$, 
or a power law with a Gaussian feature peak at $33.5^{+4.5}_{-5.5}~\msun$.
The functional form of the broken power-law mass spectrum is given by 
\begin{equation}
\frac{d\mathcal{R}_{\rm BBH}}{dM_1} \propto 
\begin{cases}
    M_1^{-\alpha_1} ~&{\rm for}~M_{\rm min}<M_1 \leq M_{\rm break},\\
    M_1^{-\alpha_2} ~&{\rm for}~M_{\rm break}<M_1<M_{\rm max},\\
    0                        ~& {\rm otherwise},
  \end{cases}
  \label{eq:bpmass}
\end{equation}
where $\alpha_1=1.58$, $\alpha_2=5.59$, $M_{\rm min}=3.96~\msun$, and $M_{\rm max}=87.14~\msun$ are adopted,
$M_{\rm break}=M_{\rm min}+bM_{\rm max}$ is the mass where there is a break in the spectral index ($b=0.43$),
and the smoothing function at $M_{\rm min}<M_1<M_{\rm min}+\delta_M$ ($\delta_M=4.83~\msun$)
is taken into account \citep{LIGO_GWTC_2020}.
With the mass spectrum, the average BH mass is calculated by
\begin{align}
\langle M_1 \rangle \equiv 
\frac{\bigints_{M_{\rm min}}^{M_{\rm max}} M_1\dfrac{d\mathcal{R}_{\rm BBH}}{dM_1}~ dM_1}
{\bigints_{M_{\rm min}}^{M_{\rm max}} \dfrac{d\mathcal{R}_{\rm BBH}}{dM_1}~dM_1}.
\label{eq:M1avez}
\end{align}
For the broken-power law mass spectrum that does not evolve with redshift,
the average mass of the primary BH is $\langle M_1 \rangle = 17.5~\msun$ and
the average total mass in a binary is $\langle M_{\rm tot,b}\rangle \equiv (1+q)\langle M_1\rangle =35~\msun[(1+q)/2]$,
where $q$ is the mass ratio of the two BHs.
In this paper, we adopt this mass spectrum as our fiducial model (see \S\ref{sec:GWB_1}).

The LIGO/Virgo observing O3a run has well constrained the mass-integrated
merger rate defined by
\begin{align}
R_{\rm BBH}(z) \equiv \int \dfrac{d\mathcal{R}_{\rm BBH}}{dM_1}dM_1.
\end{align}
The merger rate estimated from the GW events detected by the LIGO/Virgo O1+O2+O3 runs
is found to increase with redshift as $R_{\rm BBH}(z)=R_{\rm BBH,0} (1+z)^\kappa$, 
where $R_{\rm BBH,0}\simeq 19.1^{16.2}_{-9.0}~{\rm Gpc}^{-3}~{\rm yr}^{-1}$
and $\kappa = 1.8^{+9.6}_{-9.5}~(1.3^{+2.1}_{-2.1})$ for the broken power-law 
(power-law + peak) mass spectrum \citep{LIGO_GWTC_2020}.

\vspace{2mm}
\subsection{Modeling the BBH merger rate}
\label{sec:bbhrate}

The redshift-dependent BBH merger rate is given by a convolution of
the delay time distribution (DTD) $\Psi(t)$ for binary coalescences and the BBH 
formation rate $\dot{\rho}_{\rm BBH}(t)$;
\begin{equation}
R_{\rm BBH}(z) = \frac{1}{\langle M_{\rm tot,b}\rangle}\int _0^{t(z)} \dot{\rho}_{\rm BBH}(t') \Psi(t-t')dt',
\label{eq:r_psi}
\end{equation}
where $t(z)$ is the cosmic time at redshift $z$ and the average mass in a BBH
is assumed to be constant.
Here, we adopt a power-law distribution of the delay time;
\begin{equation}
\Psi (t) = \frac{\Psi_0}{t_{\rm min}} \left(\frac{t}{t_{\rm min}}\right)^{-n}~~~{\rm for}~ t_{\rm min}<t<t_{\rm max},
\label{eq:DTD}
\end{equation}
and $\Psi (t)=0$ otherwise, where $t_{\rm min(max)}$ is the minimum (maximum) merger timescale for binaries.
The normalization of $\Psi_0$ is determined so that the integration of Eq.~(\ref{eq:DTD}) 
from $t_{\rm min}$ to $t_{\rm max}$ is unity.
We consider the maximum merger time, which depends on the maximum binary separation, 
to be significantly longer than a Hubble time ($t_{\rm max}\gg t_{\rm H}$).
We here adopt $t_{\rm max}=10^3~t_{\rm H}$. 
We note that the choice of $t_{\rm max}$ is not important for $n\ga 1$, which we mainly focus on in the following discussion.
This type of the DTD is inspired by those of the GW-driven inspirals 
($n\simeq 1$; see \citealt{Piran_1992}) and other astrophysical phenomena 
related to binary mergers.
For instance, the DTD of type Ia supernovae has $n \simeq 1$ and $t_{\rm min}$ of 40 Myr to 
a few hundreds of Myr (\citealt{Maoz_2014A} and references therein), and that of short GRBs has $n\simeq 1$ and 
$t_{\rm min} \simeq 20$ Myr \citep{Wanderman_2010, Ghirlanda_2016}.
Population synthesis calculations reproduce DTDs with $n\simeq 1$ for BBH mergers
that hardly depend on the binary properties and their formation redshifts
\citep{Dominik_2012, K14,Tanikawa_2020}. 
Recently, \cite{Safarzadeh_etal_2020} discussed the effects of the delay-time nature of BBHs
on the stochastic GWB amplitude.

For the cosmic BBH formation rate, we consider two scenarios: (i) BBH formation follows the observed cosmic SFRD
\citep{Madau_Dickinson_2014} for the low-$z$ BBH population and 
(ii) BBH formation follows the SFRD given by Eq.~(\ref{eq:fid_SFRD}) for the high-$z$ BBH population.
The total stellar mass densities are $\rho_\star \simeq 5.7^{+1.7}_{-1.9}\times 10^8~\msun~{\rm Mpc}^{-3}$ for the low-$z$ population
\citep[][and references therein]{Madau_Dickinson_2014} and $\rho_\star \la 10^7~\msun~{\rm Mpc}^{-3}$ for the high-$z$ population 
(see Eq.~\ref{eq:rhostar_app}).
The cosmic BBH formation rate is given by calculating a mass fraction $f_{\rm BBH}(\equiv \dot{\rho}_{\rm BBH}/\dot{\rho}_\star)$ 
of BBHs merging within $t_{\rm max}$ to the total stellar mass.
The merging-BBH formation efficiency $f_{\rm BBH}$ is estimated as a product of the following three fractions:

\begin{enumerate}
\item The first one is the mass fraction of massive stars forming BHs in a given mass budget.
Non-rotating stars of zero-age main sequence mass $M \ga M_\bullet = 20~\msun$
are expected to leave remnant BHs via gravitational collapse at the end of their lifetime 
\citep[e.g.,][]{Spera_Mapelli_2017}.
The mass fraction is estimated for a given IMF of $dN/dM (\propto M^{-\alpha})$ by
\begin{align}
f_\bullet \equiv \dfrac{\bigints_{M_\bullet}^{M_{\star, \rm max}} M \dfrac{dN}{dM} dM}
{\bigints_{M_{\star, \rm min}}^{M_{\star, \rm max}} M \dfrac{dN}{dM} dM}.
\label{eq:fracmb}
\end{align}
For a Salpeter IMF ($\alpha =2.35$) with a mass range of $0.1(1)-100~\msun$,
we estimate $f_\bullet \simeq 0.074~(0.189)$\footnote{
For the Salpeter IMF, the average mass of massive stars with $M\geq M_\bullet$ is estimated 
as $\langle M_\star \rangle \simeq 37~\msun$, which is larger than the average mass of the primary BH.
%Due to uncertainties on stellar/binary evolution, the relation between the stellar IMF and BH mass function
%for GW sources has not been understood. 
We note that if we use $\langle M_\star \rangle$ instead of $\langle M_1\rangle =17~\msun$, the merger rate 
in Eq.~(\ref{eq:r_psi}) is reduced by a factor of $\sim 2$ and the GWB spectrum is skewed to lower frequencies, 
but the total GWB energy density is not affected (see also \S\ref{sec:gwb_th}).
This twofold difference in the merger rate can be absorbed in the uncertainty of $f_\bullet$ by changing 
$M_{\rm min}$ from $0.1~\msun$ to $1~\msun$.}. 
However, as discussed in many previous studies in literature, the stellar evolution for such 
massive stars would suffer from a significant mass loss unless they are low-metallicity stars with 
$Z\la Z_{\rm crit}\sim 0.1~\zsun$ \citep[][and references therein]{Abbott_2016_Astro} and 
the value of $f_\bullet$ would be lower with the metallicity increasing.
There are also several lines of observational evidence that the fraction of high-mass X-ray binaries 
increases with redshift and the trend would be explained by the lack of metallicity at higher redshifts
\citep{Crowther_2010,Mirabel_2011,Fragos_2013,Mirabel_2019}.
Although we do not specify the metallicity range of the high-$z$ stellar population, 
we implicitly assume $Z\sim 0.02~Z_\odot$ 
%they are supposed to be as metal-poor as $Z\sim 0.02~Z_\odot$, 
consistent with the production rate of 
ionizing radiation we adopt in \S\ref{sec:reionmodel}.
We also note that even with $Z\la  0.1~\zsun$, the remnant mass of massive stars would be 
affected by pulsation-driven winds in the main-sequence and giant phases \citep{Nakauchi_2020} and 
by (pulsational) pair-instability supernovae in the later phases \citep{Woosley_2017,Spera_Mapelli_2017}, although the mass and 
metallicity criteria for the mass-loss process depend on the nuclear burning rate of $^{12}{\rm C}(\alpha,\gamma)^{16}{\rm O}$
and the treatments of stellar convection \citep[e.g.,][]{Farmer_2020,Costa_2021}. 

\item Secondly, we assume that those massive stars that will collapse to BHs have a binary companion
at a fraction of $f_{\rm bin}\simeq 0.7$, which is consistent with the field binary fraction of O-type stars 
at the present ($f_{\rm bin}\simeq 0.69\pm 0.09$) \citep{Sana_2012}.
Local observations suggest that the mass-ratio $q$ distribution is characterized by a power-law of $q^{\beta_q}$,
where $\beta_q = -0.1\pm 0.58$ \citep{Sana_2012} and $\beta_q = 0.1\pm 0.3$ \citep{Moe_2017} over $0.3\la q \la 1$.
However, there are no observational constraints on the $q$-distribution for low-metallicity massive binaries 
that are considered to be the progenitors of BBHs. 
On the other hand, the power-law index for BBH mergers is inferred from LIGO/Virgo detections
as $\beta_q = 1.4^{+2.5}_{-1.5}$, suggesting a concentration to $q\simeq 1$.
In this paper, we assume the mass ratio to be unity $q=1$ for simplicity, which provides an upper bound of $f_{\rm BBH}$
(i.e., a massive star has a massive binary companion at a chance of $f_{\rm bin}$).

\item 
Thirdly, only a fraction $f_{\rm mrg}$ of the massive binaries end up as BBHs merging 
within $t_{\rm max}$ due to shorter initial binary separations and/or hardening process 
through binary interactions.
Assuming that \"Opik's law is applied to massive binaries, the cumulative distribution of primordial
binary separations is logarithmically flat between $a_{\rm min}\simeq 10~R_\odot$
and $a_{\rm max}\simeq 10^6~R_\odot$\footnote{
Note that the minimum separation is set so that the primary star does not fill 
its Roche lobe at the minimum separation; namely
$a_{\rm min}\simeq R_{\rm L}/0.38 \ga 8~R_\odot (R_\star/3~R_\odot)$ \citep{Eggleton_1983}.}.
Truncating the distribution at $a\simeq 1~{\rm AU}$, for which the GW merger timescale is $\simeq t_{\rm max}$,
the fraction is estimated as $f_{\rm mrg}\simeq 0.26$.
The orbital-period distribution of observed O-type stars would prefer close binaries
more than predicted by the \"Opik's law \citep{Sana_2012,Moe_2017}, suggesting a larger value of $f_{\rm mrg}$.
In such close binaries, however, their orbital evolution is likely affected by binary interactions (e.g., mass transfer, tidal effect,
and common envelope phases) before they form BBHs
and thus the processes bring large uncertainties for estimating $f_{\rm mrg}$.
Moreover, with higher metallicities ($Z\ga Z_{\rm crit}$), mass loss from a binary system makes the binary separation 
significantly wider and its merger timescale much longer than $t_{\rm max}$.

\end{enumerate}

Using the three fractions, we calculate the merging-BBH formation efficiency as 
$f_{\rm BBH} = f_\bullet f_{\rm mrg} \left(\frac{2f_{\rm bin}}{1+f_{\rm bin}}\right)$.
For the high-$z$ BBH scenario, we adopt $f_{\rm BBH}\simeq 0.018$ for our fiducial case
($f_\bullet=0.074$, $f_{\rm bin}=0.7$, and $f_{\rm mrg}=0.3$),
although a higher value of $f_\bullet$ would be expected for more top-heavy IMF of low-metallicity stars
(see \S\ref{sec:pop3} and \ref{sec:gwb_th}).
On the other hand, the merging-BBH formation efficiency for the low-$z$ population is determined so that the local merger rate 
of the low-$z$ BBHs equals the observed GW event rates of $R_{\rm BBH,0}=19.1~{\rm Gpc}^{-3}~{\rm yr}^{-1}$.
This method allows us to avoid numerous uncertainties in modeling of the metallicity effect on the stellar evolution 
and binary interaction.
The cumulative low-$z$ stellar mass density reaches $\rho_\star \simeq (0.5-1.6)\times 10^8~\msun~{\rm Mpc}^{-3}$ 
by $z\ga 2-3$ before the cosmic noon, when metal-enrichment of the universe has not proceed yet 
but low-metallicity environments with $Z<Z_{\rm crit}$ still exist.
Therefore, the merging-BBH formation efficiency for the high-$z$ population needs to be at least $\ga 20$ times higher than 
that for the low-$z$ population so that both the populations lead to a comparable GW event rate in the local universe
(assuming that the two populations follow the same mass spectrum and DTD).
%Therefore, the BBH formation efficiency for the high-$z$ population needs to be significantly higher than 
%that for the low-$z$ population so that both the populations lead to a comparable GW event rate in the local universe.
%In fact, the enhanced factor is estimated as $\simeq 1.5~\rho_{\star}(z>2)/\rho_{\star}(z>z_{\rm reion})\simeq $ .
%Therefore, if the BBH formation efficiency for the high-$z$ population is $\sim 10$ times higher than that 
%for the low-$z$ population (assuming that the two population follow the same mass spectrum and DTD), 
%the GW event rate in the local universe would be dominated by the high-$z$ BBH population.
%
We also note that if metal-poor environments are not required for BBH formation, 
the ratio of the two efficiencies is boosted up to $\simeq 300$.
This higher contrast is required because the total stellar mass for the low-$z$ population (without the metallicity condition)
is $\sim 60$ times higher than that of the high-$z$ population and a larger number of the low-$z$ BBHs can merge at 
$z\simeq 0$ with shorter coalescence timescales.
%($\rho_{\star, \rm low}/\rho_{\star, \rm high}\simeq 55$) 
%and the number ratio of BBHs that merge at $z=0$.

%%%%%%%
%    Fig. 4   %
%%%%%%%
\begin{figure*}
\begin{center}
\begin{tabular}{cc}
{\includegraphics[width=85mm]{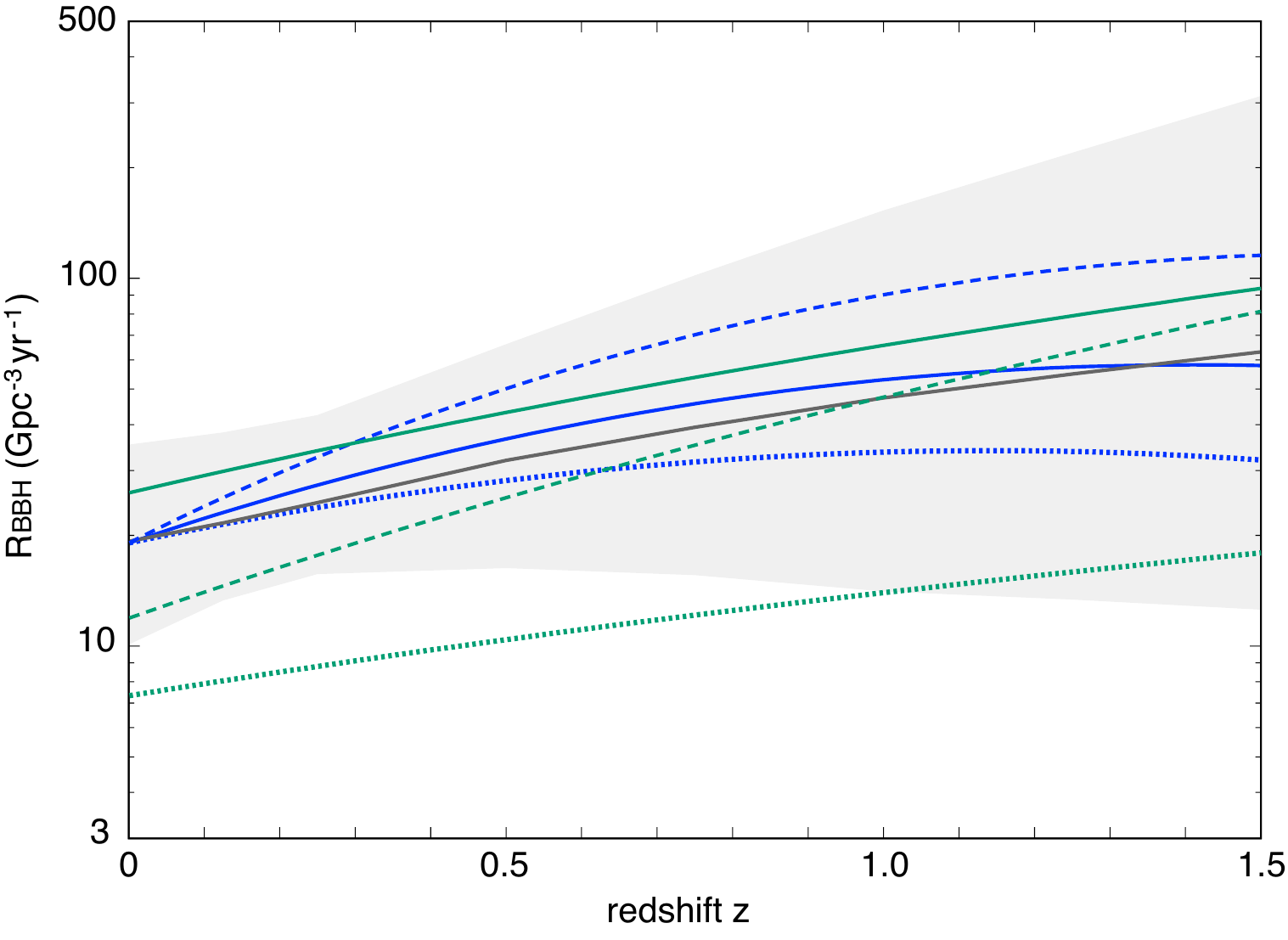}}
\hspace{5mm}
{\includegraphics[width=83mm]{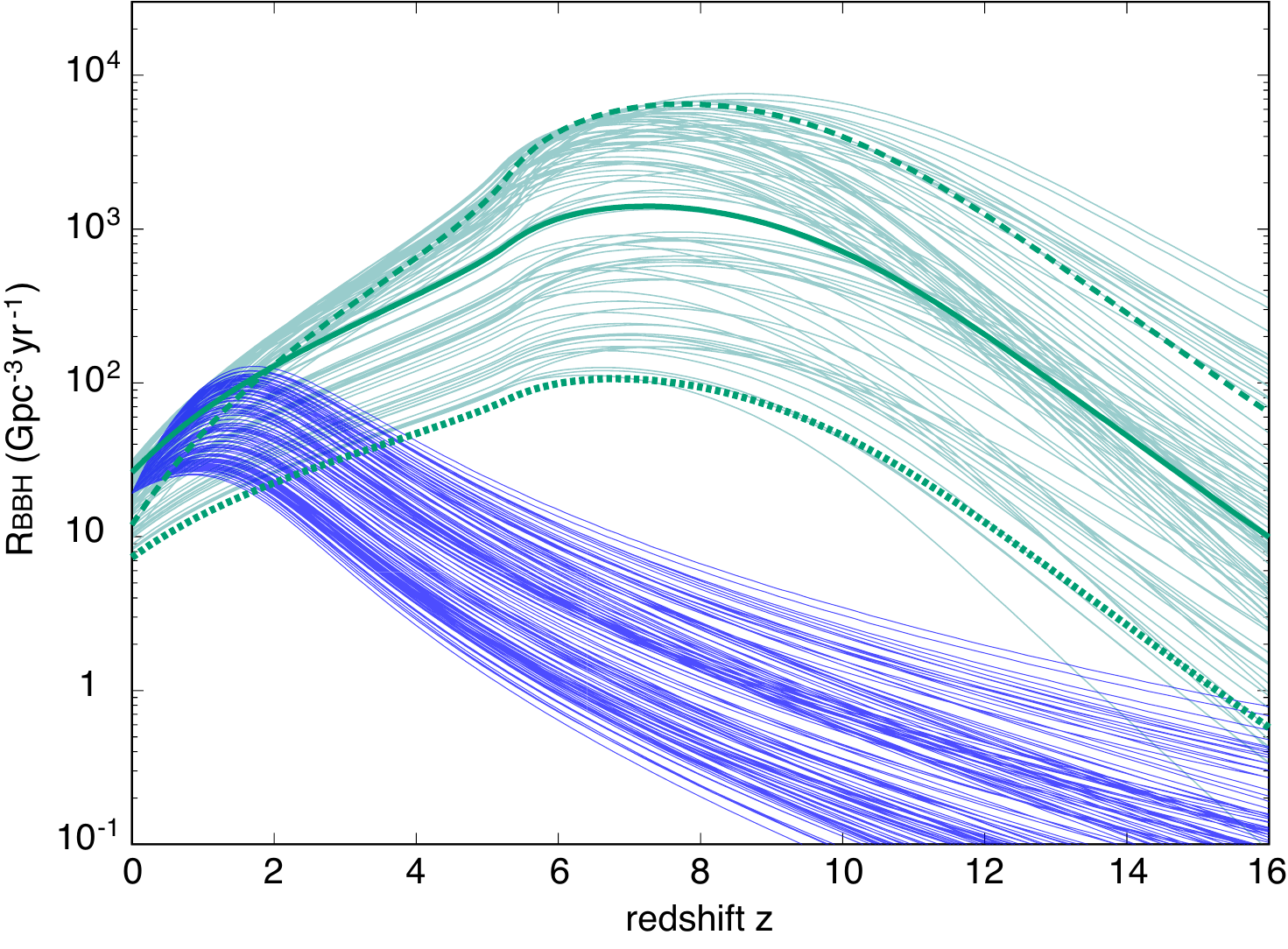}}
\end{tabular}
\caption{Redshift-dependent BBH merger rates for the low-$z$ (blue) and high-$z$ (green) BBH populations.
While the merger rate for the low-$z$ BBHs is normalized to be the observed merger rate at $z=0$,
the normalization for the high-$z$ BBHs is given by the merging-BBH formation efficiency of $f_{\rm BBH}=0.018$ 
(see \S\ref{sec:bbhrate}).
{\it Left panel}: we adopt the DTD index of $n=0.7$ (dotted), $1.0$ (solid), and $1.5$ (dashed) 
and minimal merger time of $t_{\rm min}=50$ Myr.
The redshift-dependence of all the models at lower redshifts ($z\lesssim1.5$) is overall consistent with 
that inferred from the LIGO/Virgo observing O3a run (shaded region; \citealt{LIGO_GWTC_2020}). 
{\it Right panel}: 100 different rates are generated by assuming that $n$ and $t_{\rm min}$ are distributed 
uniformly over the range of $0.7\leq n \leq 1.5$ and $10\leq  t_{\rm min}/{\rm Myr}\leq 100$.
The three cases shown in the left panel are highlighted with the green thick curves.
The merger rates for the high-$z$ BBHs rise to $R_{\rm BBH,peak}\sim 10^{3-4}~{\rm Gpc}^{-3}~{\rm yr}^{-1}$ 
at $z\sim 6-10$, which are $\sim 10-100$ times higher than those for the low-$z$ BBHs 
even though the expected BBH merger rates at $z\simeq 0$ for the both populations are comparable.
}
\label{fig:merger_rates}
\end{center}
\vspace{2mm}
\end{figure*}

It is worth giving an analytical formula of the merger rate at $z\simeq 0$ for the high-$z$ BBH population.
We approximate their SFRD as $\dot{\rho}_\star \simeq \rho_\star \delta(t-t_0)$ because their formation has 
terminated in a short duration and the detailed star formation history does not matter as long as those stars 
form at sufficiently higher redshifts; we adopt $t_0 \simeq 680~{\rm Myr}~(\ll t_{\rm H})$ 
corresponding to the cosmic time at $z\simeq z_{\rm reion}^{50\%}$.
Therefore, the merger rate is simply expressed by 
\begin{align}
& R_{\rm BBH}\simeq \frac{f_{\rm BBH}\rho_\star \Psi_0}
{\langle M_{\rm tot,b} \rangle t_{\rm min}}\left(\frac{t}{t_{\rm min}}\right)^{-n}
\label{eq:analytic}\\[3pt]
\simeq  &
\begin{cases}
\dfrac{f_{\rm BBH}\rho_\star}{\langle M_{\rm tot,b} \rangle \ln(t_{\rm max}/t_{\rm min})} 
\cdot \dfrac{1}{t} & {\rm for}~n=1,\\[10pt]
\dfrac{f_{\rm BBH}\rho_\star}{\langle M_{\rm tot,b}\rangle}\dfrac{(n-1)t_{\rm max}^n}
{t_{\rm min}t_{\rm max}^n-t_{\rm max} t_{\rm min}^n} 
\left(\dfrac{t}{t_{\rm min}}\right)^{-n} & {\rm for}~n\neq 1,\nonumber
\end{cases}
\end{align}
where $t\simeq t_{\rm H}(\gg t_0)$ is considered.
Using Eq.~(\ref{eq:analytic}), the local rate is estimated as $R_{\rm BBH}\simeq 29~{\rm Gpc}^{-3}~{\rm yr}^{-1}$
for $n=1$ ($f_{\rm BBH}=0.018$, $\rho_\star =10^7~\msun~{\rm Mpc}^{-3}$, 
$\langle M_{\rm tot,b} \rangle =35~\msun$, $t_{\rm min}=50$ Myr, $t_{\rm max} = 10^3~t_{\rm H}$).

\vspace{2mm}
\subsection{Redshift-dependent merger rates of \\the two BBH populations}

In Fig.~\ref{fig:merger_rates}, we show the redshift-dependent BBH merger rates 
for the low-$z$ (blue curves) and high-$z$ (green curves) BBH populations.
In the left panel, each curve is generated by setting $t_{\rm min}=50$ Myr and $n=0.7$ (dotted), 
$1.0$ (solid), and $1.5$ (dashed).
The redshift-dependence of all the models at lower redshifts ($z\lesssim1.5$) is overall consistent with 
that inferred from the LIGO/Virgo O3a run (\citealt{LIGO_GWTC_2020}; 90\% credible intervals 
shown by the gray shaded band).
The merger rates for the high-$z$ BBHs are well described by Eq.~(\ref{eq:analytic})
and could explain most GW events observed at $z\simeq 0$ in terms of the rate,
only if the merging-BBH formation efficiency is as high as $f_{\rm BBH}\simeq 2\%$
(note that the merger rate scales with the value of $f_{\rm BBH}$).

In the right panel, we show the BBH merger rates for the two populations
extending the redshift range up to $z=16$.
For each model, we generate 100 different rates by assuming that the power-law index $n$ 
and the minimum merger time are distributed uniformly over the range of $0.7 \leq n \leq 1.5$ 
and $10 \leq t_{\rm min}/{\rm Myr} \leq 100$
(the three cases with $t_{\rm min}=50$ Myr shown in the left panel are highlighted with green thick curves).
For the low-$z$ population, the merger rates have peaks of $R_{\rm BBH,peak} \sim 30-100~{\rm Gpc}^{-3}~{\rm yr}^{-1}$ 
at the epoch when the cosmic star formation rate is the highest, and decreases toward higher redshifts.
In contrast, for the high-$z$ population, a vast majority of the BBHs merge in the early universe at $z\simeq 6-10$ 
and a small fraction of them (binaries with wider orbital separations at birth) merge 
within the LIGO/Virgo detection horizon.
For the high-$z$ BBH population, the shape of the merger rate depends on the DTD index more sensitively.
For the canonical case ($n=1$; solid), the merger rate increases to $R_{\rm BBH,peak} 
\sim 10^3~{\rm Gpc}^{-3}~{\rm yr}^{-1}$ at $z\sim 6-10$, which is $>10$ times higher than for the low-$z$ BBHs,
even though the expected local rate is similar to that for the low-$z$ BBH population.
With the larger DTD indices ($n=1.5$; dashed), most BBHs merge at higher redshifts at a peak rate of 
$\sim 6\times 10^3~{\rm Gpc}^{-3}~{\rm yr}^{-1}$, but the rate quickly decays toward $z\simeq 0$
because the total mass of BBHs is fixed.
With the smaller DTD indices ($n=0.7$; dotted), most BBHs do not merger within a Hubble time and 
thus both the peak and local rate are significantly lowered.

\begin{figure}
\begin{center}
\includegraphics[width=85mm]{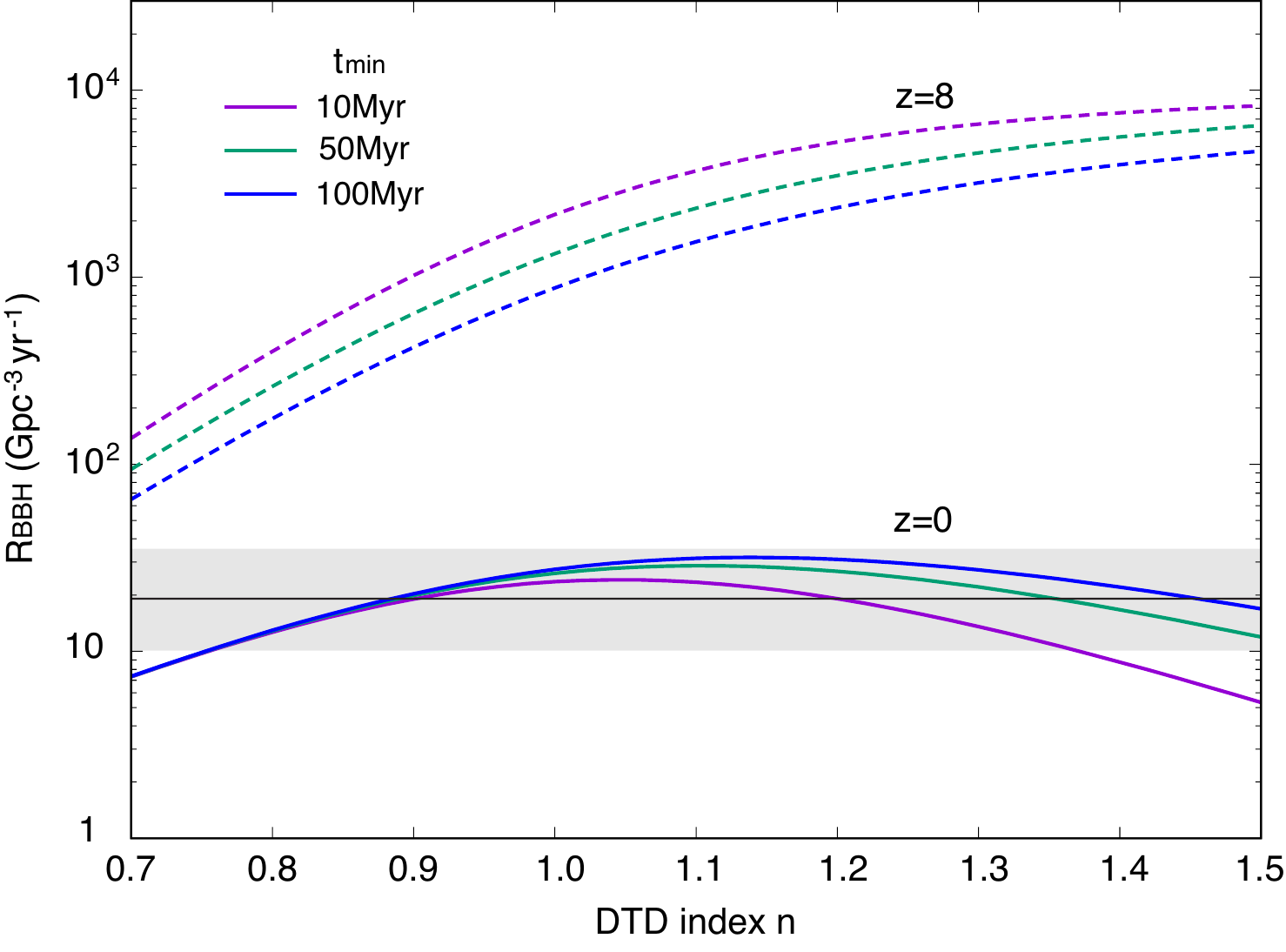}
\vspace{-2mm}
\caption{Summary of the dependence of the merger rates for the high-$z$ BBHs at $z=0$ (solid) and 
$z=8$ (dashed) on the DTD index $n$.
Each curve corresponds to the case with different minimum merger time:
$t_{\rm min}=10$ (purple), $50$ (green), and $100$ Myr (blue).
}
\label{fig:dist_param}
\end{center}
\end{figure}

In Fig. \ref{fig:dist_param}, we summarize the dependence of the high-$z$ BBH merger rates 
on the DTD index $n$.
Here, we focus on the merger rate at $z=0$ (solid curves) and $z=8$ (dashed), when the rate is maximized.
Each curve corresponds to the case with different minimum merger time:
$t_{\rm min}=10$ (purple), $50$ (green), and $10$ Myr (blue).
As also seen in Fig.~\ref{fig:merger_rates}, the local merger rate is maximized at $n\simeq 1$
because a good fraction of BBHs formed at $z>z_{\rm reion}$ merge within a Hubble timescale.
With a shorter $t_{\rm min}$, the local rate decreases but the peak rate at $z\sim 8$ increase,
reflecting the conservation of the total BBH mass budget.
The local rate depends on $t_{\rm min}$ only when the DTD index is larger than unity,
i.e., the normalization of the DTD determined by the choice of $t_{\rm min}$.
Overall, the merger rates for a wide range of the DTD parameters explain the local GW event rate
inferred from the LIGO/Virgo O3a observing run (gray region).
The peak merger rate increases with the DTD index but approaches 
$R_{\rm BBH}\sim 10^{4}~{\rm Gpc}^{-3}~{\rm yr}^{-1}$ at $n\ga 1.3$.
We note that the apparent maximum rate corresponds to the case where
all the BBHs immediately merger at birth; 
$R_{\rm BBH}\sim f_{\rm BBH}\dot{\rho}_\star / \langle M_{\rm tot,b}\rangle \simeq 
1.6\times 10^{4}~{\rm Gpc}^{-3}~{\rm yr}^{-1}$.

%%%%%%%
%    Fig. 5   %
%%%%%%%

Finally, we briefly mention the merger rate of BBHs originating from PopIII stars.
As discussed in \S\ref{sec:pop3} (see Fig.~\ref{fig:app3}), the upper limit of the mass density for
PopIII stars is limited below $\rho_{\star,\rm III} \la 2\times 10^5~\msun~{\rm Mpc}^{-3}$,
which is $\sim 2\%$ of that for the normal PopII stars.
Therefore, even if PopIII BBHs follow the DTD with $n\simeq 1.0$,
the merger rate of PopIII BBHs at $z\simeq 0$ would be as small as 
$R_{\rm BBH,III}\simeq 0.5~(f_{\rm BBH}/0.018)(\langle M_{\rm tot,b}\rangle/35~\msun)^{-1}~{\rm Gpc}^{-3}~{\rm yr}^{-1}$.
This indicates that they could contribute to the local GW events, {\it only if
$\ga 40\%$ of all the mass in PopIII stars would be converted into BBHs merging within a Hubble time}.
Recently, \cite{Kinugawa_2021} claimed that BBHs originating from PopIII remnants could explain 
the local GW event rate at $M_1\ga 30~\msun$, which is responsible for $\sim 10~{\rm Gpc}^{-3}~{\rm yr}^{-1}$
and requires $f_{\rm BBH}\simeq 0.5$ for $\langle M_{\rm tot,b}\rangle =50~\msun$ (note that they adopt $q=0.7$).
Such a high merging-BBH formation efficiency could be provided for a top-heavy IMF (e.g., a flat IMF with a mass range of $10-100~\msun$;
$f_\bullet \simeq 0.96$), a high binary fraction $f_{\rm bin}\simeq 1$, and
$f_{\rm mrg}\ga 0.5$ (e.g., the distribution of primordial binary separations prefer close binaries;
see also \citealt{Inayoshi_2017}).
Dynamical capture of BHs in dense metal-free clusters would also form tightly bound BBHs 
\citep{Liu_Bromm_GW_2020}.

\vspace{2mm}
%%%%%%
% Sec. 3 %
%%%%%%
\section{Gravitational wave background}\label{sec:GWB}
We next calculate the spectrum of a GWB produced from BBHs that 
merge at the rates shown in Fig. \ref{fig:merger_rates};
\begin{equation}
\rho_{\rm c}c^2\Omega_{\rm gw}(f)=
\int _{z_{\rm min}}^\infty
\int_{M_{\rm min}}^{M_{\rm max}}
\frac{d\mathcal{R}_{\rm BBH}}{dM_1}
\left(f_r\frac{dE_{\rm gw}}{df_r}\right) \frac{dt}{dz}~\frac{dM_1 dz}{1+z},
\label{eq:omega_gw_2}
\end{equation}
\citep{Phinney_2001},
where $f$ and $f_r$ are the GW frequencies observed at $z=0$ and in
the source's rest frame, i.e., $f_r=f(1+z)$, and $\rho_{\rm c}$ is the
critical density of the universe.  
We set the minimum redshift to $z_{\rm min}=0.28$, the detection horizon of LIGO/Virgo\footnote{
Given the GW sensitivity curve, the size of the observational horizon for a BBH merger depends on the masses of the two BHs.
For simplicity, we adopt one single value of the redshift within which BBHs are individually detected.
However, the choice weakly affects the estimation of a GWB
only for the low-$z$ BBH population if $z_{\rm min}>0.3$ is set.
In this sense, the calculated GWB amplitude for the low-$z$ BBH population corresponds to an upper limit.
}.
The GW spectrum from a coalescing BBH is given by
\begin{equation}
\frac{dE_{\rm gw}}{df_r}=\frac{(\pi G)^{2/3}M_{\rm c}^{5/3}}{3f_r^{1/3}}
\mathscr{F_{\rm PN}}
\label{eq:omega_gw_1}
\end{equation}
where $E_{\rm gw}$ is the energy emitted in GWs, $M_{\rm c} \equiv (M_1M_2)^{3/5}/(M_1+M_2)^{1/5}$ 
is the chirp mass, $M_2$ is the secondary mass, and $\mathscr{F_{\rm PN}}$ is the Post-Newtonian 
correction factor \citep{Ajith_2011}.
We here consider merger events of equal-mass binaries to be consistent with previous works
\citep{LIGO_back_2016,LIGO_GWTC_2019}, which differ from the conditional mass-ratio $q$ 
distribution of $q^{\beta_q}$ ($\beta_q = 1.4^{+2.5}_{-1.5}$ for the broken power-law mass spectrum) 
inferred from the observed merger events \citep{LIGO_GWTC_2020}.
We note that assuming $q=0.7$, the GWB amplitude shown below is reduced at most by a factor
of $\simeq 1.3$ ($\simeq 80\%$) at $f<100~{\rm Hz}$.
This level of small reduction would be absorbed in the uncertainties of $\beta_q$ and other model parameters
characterizing the primary mass function.
We also assume that the orbits of BBHs that contribute to a GWB are circularized by the time 
they move into the LIGO/Virgo band and thus the GWB spectrum in the inspiral phase scales 
with frequency as $\Omega_{\rm gw}(f)\propto f^{2/3}$.
While binary-single interactions can produce high-eccentricity events, they are likely to constitute 
a significant fraction of all events only in the AGN disk models \citep[e.g.,][]{Tagawa_2020,Samsing_2020}.

%%%%%%%
%    Fig. 6   %
%%%%%%%
\begin{figure}
\begin{center}
{\includegraphics[width=85mm]{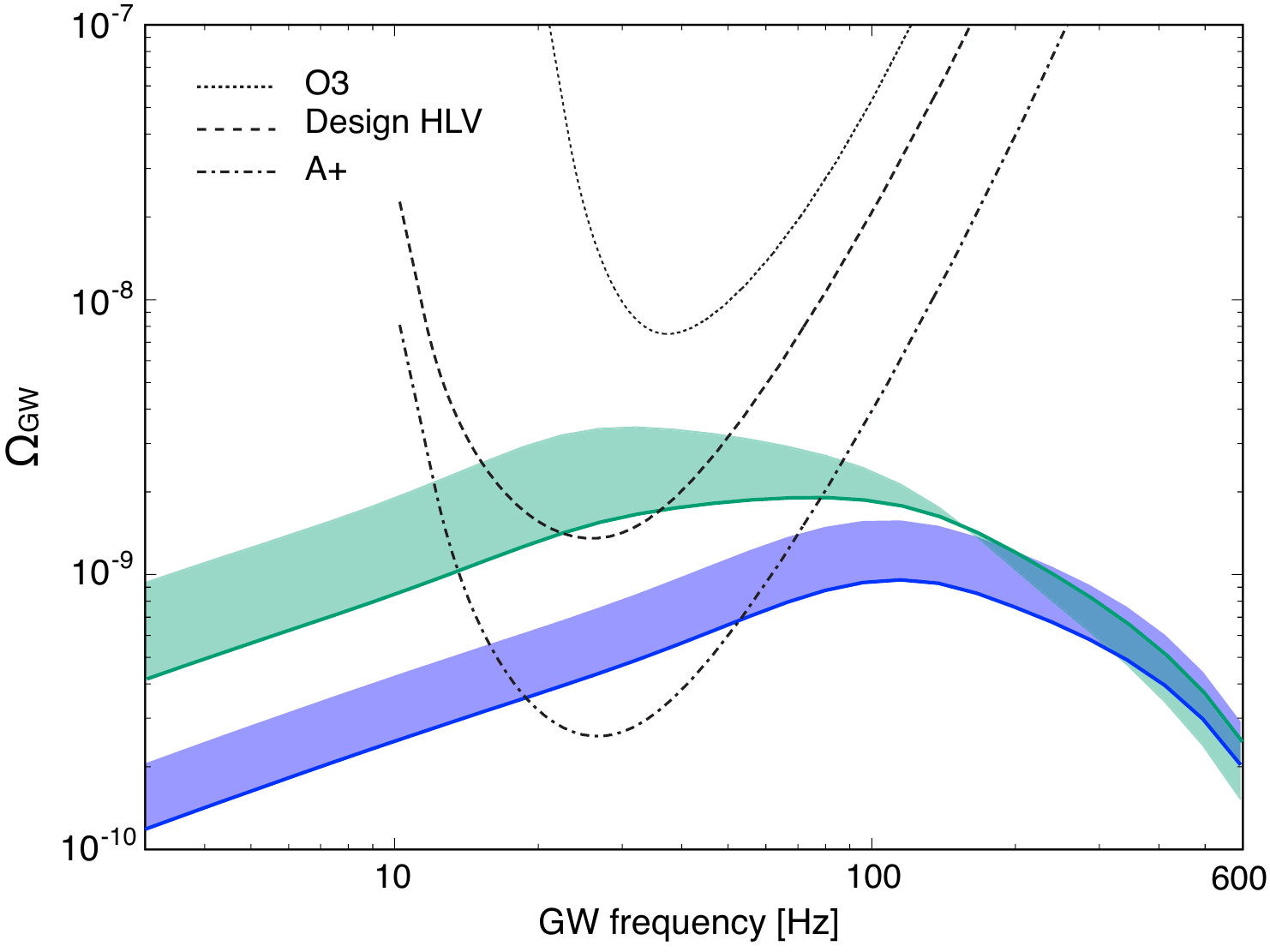}}
\caption{The stochastic GWB spectra produced by the low-$z$ (blue) and high-$z$ (green) BBH populations
that follow the merger rates shown in Fig.~\ref{fig:merger_rates}.
For each case, the shaded region shows the expected GWB amplitude for different DTD parameters;
namely $1.0\leq n\leq 1.5$ and $t_{\rm min}=50~{\rm Myr}$.
The three sensitivity curves of the O3 run (dotted), the HLV design (dashed), and the envisioned A$^+$ (dot-dashed) are shown.
}
\label{fig:GWB}
\end{center}
\end{figure}

\subsection{The mass function of BBH mergers consistent with locally detected GW sources}\label{sec:GWB_1}

First, we consider BBH mergers whose mass function follows the broken power law
provided by the most updated samples of locally detected GW sources (see Eq.~\ref{eq:bpmass}).
We assume that the mass function shape does not evolve with redshift,
while the mass-integrated merger rate evolves as shown in \S\ref{sec:bbhrate}.

In Fig.~\ref{fig:GWB}, we present the stochastic GWB spectra for the low-$z$ and 
high-$z$ BBH populations, along with the three sensitivity curves of
the O3 run (dotted), the HLV design (dashed), and A$^+$ (dot-dashed)\footnote{HLV stands for 
LIGO-Hanford, LIGO-Livingston, and Virgo; https://dcc.ligo.org/LIGO-G2001287/public}.
The BBH merger rate for each population is shown in Fig.~\ref{fig:merger_rates}.
The shaded regions show the expected GWB amplitude for different DTD indices at $1.0\leq n\leq 1.5$
(the solid curves for $n=1.0$).
The minimum merger time is set to $t_{\rm min}=50~{\rm Myr}$ for the two populations
since the GWB amplitude hardly depends on the choice as long as $t_{\rm min}$ is much less than 
$\sim 10$ Gyr.

For the low-$z$ BBHs, regardless of the model uncertainties, the spectral shape of the GWB
is characterized by a well-known (lowest Newtonian order) power-law of $\Omega_{\rm gw}(f) \propto f^{2/3}$ at $f<100~{\rm Hz}$ 
and peaks at higher frequencies \citep[for comparison, see][]{LIGO_back_2016}. 
The GWB amplitude is as low as $\Omega_{\rm gw}\simeq 4.14_{-1.45}^{+1.87}\times 10^{-10}$ at
$f=25~{\rm Hz}$, where the LIGO/Virgo detectors are the most sensitive.
As already pointed out in \cite{LIGO_GWB_2021}, the weak GWB signal is not detectable 
at the LIGO/Virgo design sensitivity, but requires the envisioned A$^+$ sensitivity to be detected.

For the high-$z$ BBHs, the GWB amplitude is as high as 
$\Omega_{\rm gw}\simeq 1.48_{-1.27}^{+1.80}\times 10^{-9}$ at $f=25~{\rm Hz}$.
The GWB spectrum is significantly flatter at $f\ga 20-30~{\rm Hz}$ from the value 
of $2/3$ and peaks inside the frequency window of the LIGO/Virgo observations.
This characteristic spectral shape predicted by \cite{Inayoshi_2016} still holds in this modeling where 
the most updated properties of merging BBHs provided in the GWTC-2 catalog is used.
Note that the detailed properties of the spectral flattening depends on model parameters
as seen in previous studies \citep{Inayoshi_2016, Perigois_2020}.
Even if the constraint from cosmic reionization is taken into account, the GWB signal is still detectable 
at the HLV design sensitivity.
Moreover, if the DTD index is larger than unity, the unique feature of the GWB spectrum can be detected 
with the HLV design sensitivity.
In addition, the detection of this level of GWB  would indicate a major contribution by the high-redshift BBH 
population to the local GW events.

%%%%%%%
%    Fig. 7   %
%%%%%%%
\begin{figure}
\begin{center}
{\includegraphics[width=85mm]{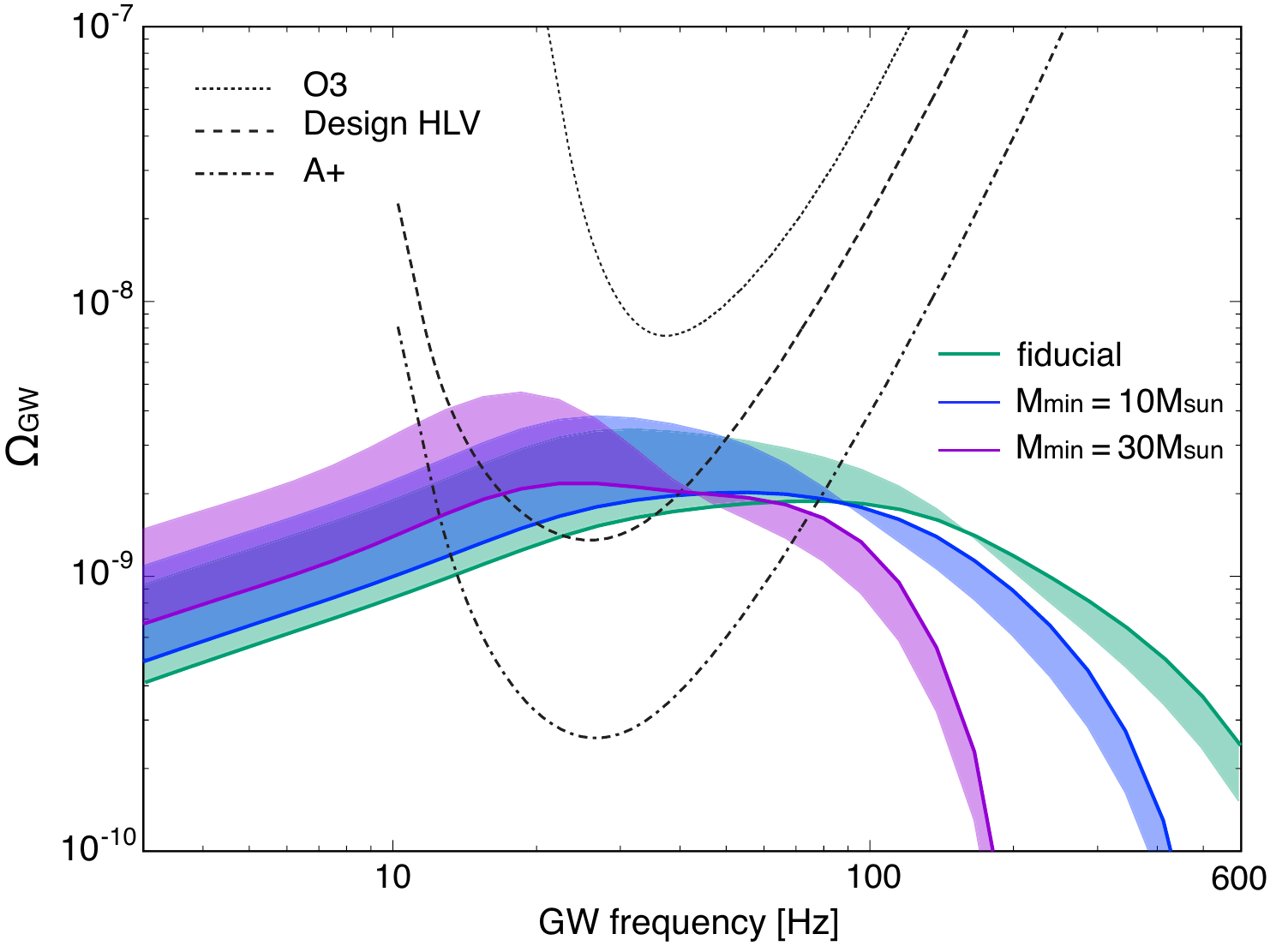}}
\caption{Same as in Fig.~\ref{fig:GWB} but with different top-heavy merger mass functions with 
$M_{\rm min}=3.96$ (green, fiducial case), $10$ (blue), and $30~\msun$ (purple).
The shaded region presents the expected GWB signal in each model with a DTD index between
$1.0\leq n \leq 1.5$ (the solid curve for $n=1.0$).
When the high-$z$ BBHs follow more top-heavy mass functions than in the local universe,
the spectral shape is skewed toward lower frequencies and the characteristic flattening is detectable at 
the HLV design sensitivity.
}
\label{fig:GWB_mass_evo}
\end{center}
\end{figure}

The existence of such individually undetectable BBH mergers beyond the detection horizon also
serve as a source of GW events that can be gravitationally lensed by the foreground structures 
\citep[e.g.,][]{Dai_2017,Oguri_2018,Contigiani_2020,Mukherjee_2021}.
However, \cite{Buscicchio_2020} recently showed that even assuming a merger rate at $z>1$ high enough 
to produce a GWB detectable at the HLV design sensitivity, the lensing probability for individually detected BBH mergers 
is as small as $\la 10^{-3}$ over 2 years of operation.
Therefore, if the high-$z$ BBHs contribute to the production of a GWB at the predicted level, 
we will be able to detect the GWB before a lensed GW source is detected.

\subsection{The upper bound of GWBs produced from high-$z$ BBHs with more top-heavy mass function}
\label{sec:gwb_th}

As an alternative model, we consider a high-$z$ BBH population that follows
a mass function more top-heavy than the broken power-law one adopted in the fiducial model. 
This is motivated by the absence of high-mass BBH detections at low redshifts indicating
that the astrophysical BBH mass distribution evolves and/or the largest BBHs only merge at 
high redshifts \citep{Fishbach_2021}.
Moreover, a top-heavy mass function is expected from 
cosmological simulations of high-$z$ star formation \citep[e.g.,][]{Hirano_2014},
BBH formation channels \citep{K14, Inayoshi_2017}, and possible subsequent growth processes 
via gas accretion in protogalaxies and/or disks in active galactic nuclei
(\citealt{Tagawa_2020,Safarzadeh_2020}; see also \citealt{Inayoshi_HE_2016}).

In Fig.~\ref{fig:GWB_mass_evo}, we present the GWB amplitudes for three high-$z$ BBH populations
whose merger mass function is given by the broken power-law function with 
$M_{\rm min}=3.96~\msun$ (green; fiducial case), $M_{\rm min}=10~\msun$ (purple), and $M_{\rm min}=30~\msun$ (blue).
For the top-heavy models, the average mass of the primary BH is $\langle M_1\rangle = 26$ and $47~\msun$, 
respectively, which are used for estimating the merger rate (see Eq.~\ref{eq:r_psi}).
The shaded region presents the expected GWB signal in each model, 
associated with the possible range of the DTD index; 
the solid curve is for $n=1.0$ and the highest value at lower frequencies is for $n=1.5$.
Note that for the top-heavy models, the contribution from BBHs with $3.96~\msun < M_1 <M_{\rm min}$ 
to the GWB is not included in Fig. \ref{fig:GWB_mass_evo}.

With the higher minimum mass, the peak frequency of the GWB moves to a lower value
and thus the flattening of the spectrum becomes more prominent compared to the fiducial case (green).
The peculiar spectral indices are substantially lower than the canonical value of $\sim 2/3$ 
expected from lower-$z$ and low-mass BBH mergers.
Even with the constraint from cosmic reionization, the expected GWBs for the two top-heavy models 
are as strong as $\Omega_{\rm gw}\simeq 3.8\times 10^{-9}$ at $f=25$ Hz for $M_{\rm min}=10~\msun$
and $\Omega_{\rm gw}\simeq 4.5\times 10^{-9}$ at $f=20$ Hz for $M_{\rm min}=30~\msun$,
which are well above the detection thresholds with the HLV design sensitivity.
A detection of such a unique spectrum with the design sensitivity would allow us to extract information on 
a top-heavy-like mass function of the high-$z$ BBH merger population.

It is worth providing an analytical expression of the GWB upper bound
constrained by the history of cosmic reionization.
Here, we consider the total GWB energy density calculated with
\begin{align}
\mathcal{E}_{\rm GW}
&\equiv \int_0^\infty \rho_{\rm c}c^2\Omega_{\rm gw}(f)\frac{df}{f},
\label{eq:total_GWB}
\\[5pt]
&=
\eta_{\rm gw} c^2 
\int _{z_{\rm min}}^\infty
\left[
\int _{M_{\rm min}}^{M_{\rm max}}
qM_1 
\frac{d\mathcal{R}_{\rm BBH}}{dM_1}
dM_1
\right]
\frac{dt}{dz} \frac{dz}{1+z},\nonumber
\end{align}
where the GW radiative efficiency is approximated as a constant value of
$\eta_{\rm gw} \simeq 0.1$, which is valid for $q>1/3$.
Using Eqs.~(\ref{eq:r_psi}) and (\ref{eq:analytic}), the above equation is approximated as
\begin{align}
\frac{\mathcal{E}_{\rm GW}}{\rho_c c^2}
&\simeq 
\frac{\eta_{\rm gw}f_{\rm BBH} \rho_\star}{\rho _c} \cdot \frac{q}{1+q}~\mathcal{I}_n,
\end{align}
where
\begin{equation}
\mathcal{I}_n \equiv \int_{t_0+t_{\rm min}}^{t_{\rm H}} \frac{\Psi (t-t_0)}{1+z}dt,
\end{equation}
which is numerically calculated as $\mathcal{I}_{1.5}\simeq 0.164$, $\mathcal{I}_{1.0}\simeq 0.110$,
and $\mathcal{I}_{0.7}\simeq 0.031$.
Here, the SFRD is approximated $\dot{\rho}_\star \simeq \rho_\star \delta(t-t_0)$, 
$t_0~(\simeq 680~{\rm Myr})$ corresponds to the cosmic time at $z\simeq z_{\rm reion}^{50\%}$,
and the typical mass ratio does not evolve significantly.
In conclusion, we obtain the upper bound on the total GWB energy density
\begin{align}
\frac{\mathcal{E}_{\rm GW}}{\rho_c c^2}
&\simeq
8.0\times 10^{-9}
\left(\frac{2q}{1+q}\right)
\left(\frac{\eta_{\rm gw}}{0.1}\right)
\left(\frac{\mathcal{I}_n}{0.1}\right)
\nonumber\\
&~~\times
\left(\frac{f_{\rm BBH}}{0.02}\right)
\left(\frac{\rho_\star}{10^7~\msun~{\rm Mpc}^{-3}}\right).
\end{align}
We note that the total GWB energy density is independent of the merger mass function.
Depending on the GWB spectral shape, which {\it does} depend on the mass function of BBH mergers,
a fraction of the total GWB energy is distributed in the frequency band where the ground-based 
GW detectors are sensitive.

%%%%%%%
%    Fig. 8   %
%%%%%%%
\begin{figure}
\begin{center}
{\includegraphics[width=84mm]{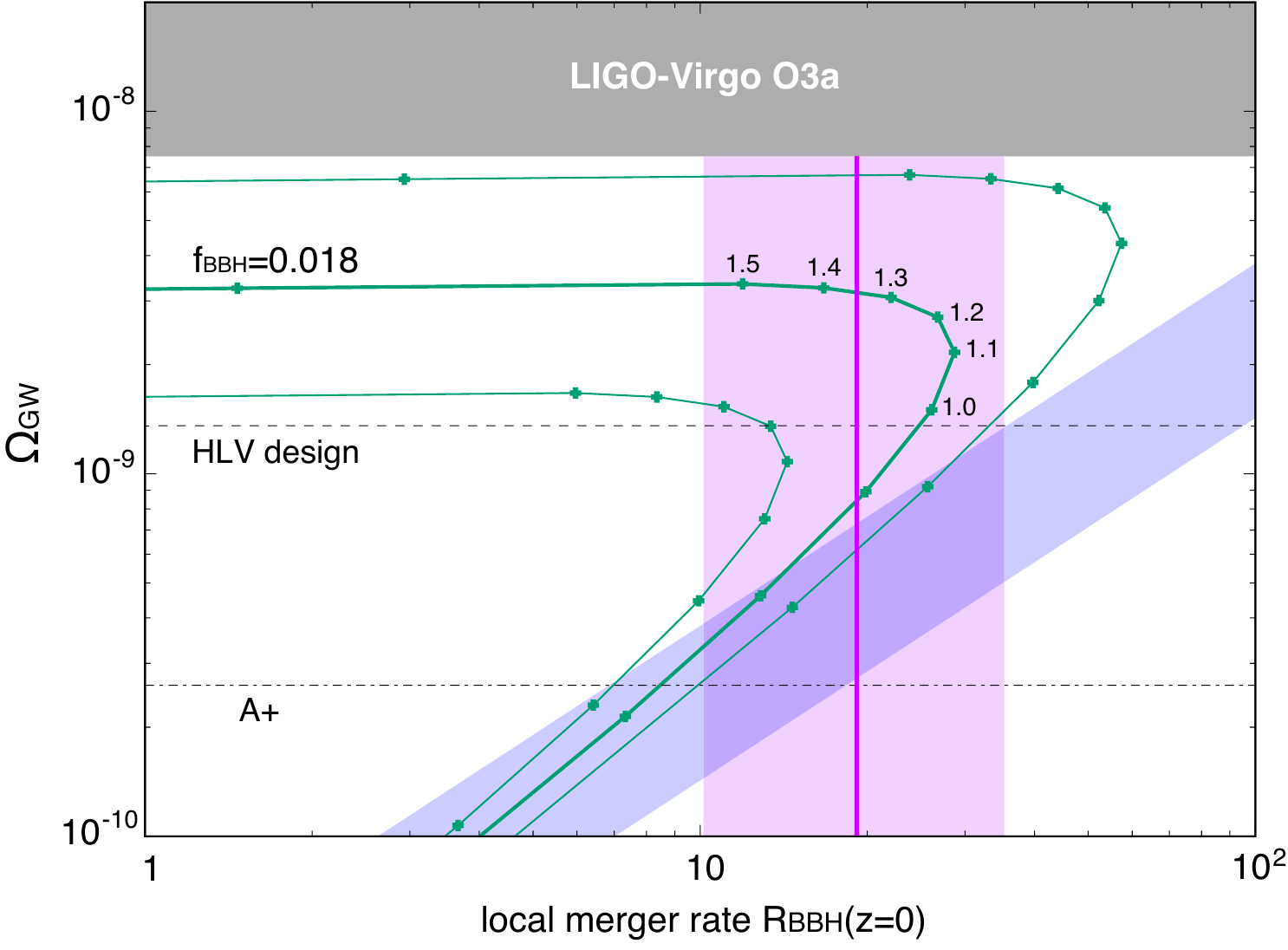}}
\caption{The relation of the GWB amplitude at $f=25$ Hz and the local BBH merger rate
for the low-$z$ (blue region; $0.7\leq n \leq 1.5$) and high-$z$ BBH (green curves; $f_{\rm BBH}/0.018 = 0.5$, $1.0$, and $2.0$) population.
The GWB amplitude for the high-$z$ BBHs increases with the DTD index (denoted by the numbers in the figure), 
while the local merger rate is maximized for $n\sim 1.0-1.2$.
In the fiducial case ($f_{\rm BBH}=0.018$; thick curve), the GWB is detectable at the design sensitivity
when the DTD index is in $1.0\la n \la 1.5$, where the expected local merger rate agrees with the observed GW event rates
(magenta region; $R_{\rm BBH,0}=19.1^{+16.2}_{-9.0}~{\rm Gpc}^{-3}~{\rm yr}^{-1}$).
The current upper limit obtained from the LIGO-Virgo O3a observing run
\citep{LIGO_GWB_2021} gives a constraint of $f_{\rm BBH}\la 0.04$.
}
\label{fig:RBBH_GWB}
%\vspace{2mm}
\end{center}
\end{figure}

\vspace{2mm}
\subsection{The relation between $\Omega_{\rm gw}$, $R_{\rm BBH,0}$, and reionization parameters}

In Fig.~\ref{fig:RBBH_GWB}, we summarize the relation of the GWB amplitude at $f=25$ Hz and 
the local BBH merger rate for each population with different model parameters.
For the low-$z$ BBH population (blue region), the expected GWB amplitude is assumed to be proportional to
the local merger rate.
For a given local merger rate, the GWB amplitude increases with the DTD index ($0.7\leq n\leq 1.5$),
but it is not detectable at the HLV design sensitivity.
For the high-$z$ BBH population (green curves), the GWB amplitude increases with the DTD index 
(denoted by the numbers in the figure), while the local merger rate is maximized for $n\sim 1.0-1.2$,
where the distribution of BBH mergers is spread logarithmically in time and thus a good fraction of BBH mergers
occur at $z\simeq 0$.
As shown in Fig.~\ref{fig:dist_param}, the local merger rate decreases for smaller and larger DTD indices 
because most mergers will be pushed into the future ($n<1$) or occurred well before $z\sim 0$ ($n>1$).
In the fiducial case ($f_{\rm BBH}=0.018$; thick curve), the GWB is detectable at the design sensitivity
when the DTD index is in $1.0\la n \la 1.5$, where the expected local merger rate agrees with the observed GW event rates
(magenta region; $R_{\rm BBH,0}=19.1^{+16.2}_{-9.0}~{\rm Gpc}^{-3}~{\rm yr}^{-1}$).
Therefore, once this level of GWB will be detected in the O5 observing run, 
this would indicate a major contribution of the high-z BBH population to the local GW events.
If the merging-BBH formation efficiency for the high-$z$ population is substantially less than $\sim 1\%$, 
the high-$z$ BBH population neither produces a GWB detectable at the design sensitivity
nor explains the local merger rate.
In this case, the low-$z$ BBH population dominates the local event rate and a GWB owing to the low-$z$ BBH population
would be detected at the envisioned A$^+$ sensitivity.
Additionally, the current upper limit of the stochastic GWB obtained from the LIGO-Virgo O3a observing run
\citep{LIGO_GWB_2021} gives a constraint of $f_{\rm BBH}\la 0.04$.

The constraint on $f_{\rm BBH}$ from the GWB detection would also be expressed as
the relation between the number of BBHs ($q=1$ is assumed) and ionizing photons produced from
the same stellar mass budget.
From Fig.~(\ref{fig:RBBH_GWB}), we obtain 
\begin{align}
\frac{\#{\rm BBH}}{\#{\rm photon}} &= 
\frac{f_{\rm bin}}{1+f_{\rm bin}}\frac{m_{\rm p}f_\bullet}{\eta_{\rm ion}\langle M_\bullet \rangle},\\[10pt]
 \simeq 
3.2 \times & 10^{-64}
\left(\frac{\langle M_\bullet \rangle}{20~\msun}\right)^{-1}
\left(\frac{f_{\rm esc}}{0.1}\right)
\left(\frac{\Omega_{\rm gw,25Hz}}{1.5\times 10^{-9}}\right)
\nonumber
\end{align}
where $\langle M_\bullet \rangle$ is the average BH mass for a given IMF, 
other parameters are fixed to their fiducial values, and the DTD index is set to $n=1.0$.
Note that we here neglect the extra numerical factor of $(f_{\rm esc}\eta_{\rm ion}/400)^{0.2}$.
Therefore, a detection of the GWB at the HLV design sensitivity ($\Omega_{\rm gw}\ga 1.4\times 10^{-9}$ at $f=25$ Hz) 
indicates the existence of a high-$z$ stellar population that forms a few BBHs per $\sim 10^{64}$ ionizing photons\footnote{
This efficiency of \#BBH/\#photon shown here ($q=1$ is assumed) is $\sim 10$ times higher than that given in 
Fig.~2 of \cite{Inayoshi_2016},
where the binary mass ratio follows a flat distribution.}.
This would give us an insight on the properties of BBH's stellar progenitors (e.g., IMF and metallicity).

%\vspace{2mm}
\subsection{Primordial binary BHs}

Finally, we briefly discuss a GWB produced by BBHs whose formation rate does not necessary 
follow the cosmic star formation history (e.g., primordial BBH population; 
see a recent review by \citealt{Carr_2020}).
The time dependence of the merger rate is calculated as $R_{\rm BBH} \propto t^{-34/37}$ 
at $z\ll 1000$ (\citealt{Nakamura_1997} and \citealt{Sasaki_2016}),
and the expected GWB amplitude ($f=25~{\rm Hz}$) owing to primordial BBHs is as weak as
$\Omega_{\rm gw,PBH}\la 10^{-9}$, which is below the HLV design sensitivity,
even assuming that all the GW events locally observed originate from the primordial BBH population.
This upper bound corresponds to the case where PBHs constitute a fraction of dark matter; 
namely $\Omega_{\rm PBH}/\Omega_{\rm m}\simeq 2\times 10^{-3}$ \citep[see more arguments in a review paper by][]{Sasaki_2018}.

\vspace{5mm}
\section{Summary and discussion}\label{sec:summary}

In this paper, we consider the gravitational wave background (GWB) produced by 
binary black hole (BBH) mergers originating from 
the high-$z$ universe at the cosmic dawn.
Since overproduction of ionizing photons from stellar progenitors of those BBHs is constrained by
the {\it Planck} measured optical depth of the universe to electron scattering,
the total stellar mass formed during the epoch of reionization has an upper bound.
Using a semi-analytical model of the reionization history, we quantify the critical stellar mass density 
for a metal-enriched PopII stellar population that lead reionization dominantly as
$\rho_\star \simeq 10^7~\msun~{\rm Mpc}^{-3}$ (see Eq.~\ref{eq:rhostar_app}).
This value is lowered if other rarer but more intense radiation sources (e.g., metal-free PopIII stars) 
could contribute to reionization.

Under this constraint from the reionization history, the merger rate for the high-$z$ BBH population
becomes as high as $R_{\rm BBH}\simeq 5-30~{\rm Gpc}^{-3}~{\rm yr}^{-1}$
at $z\simeq 0$ for a wide range of the parameters of the delay-time distribution (DTD) for BBH coalescences,
where the merging-BBH formation efficiency is assumed to be as high as $f_{\rm BBH}(\equiv \dot{\rho}_{\rm BBH}/\dot{\rho}_{\star}) \simeq 0.02$.
Since a vast majority of the BBHs merge in the early universe, the merger rate increases to 
$R_{\rm BBH}\simeq 10^{3-4}~{\rm Gpc}^{-3}~{\rm yr}^{-1}$ at $z\simeq 6-10$ for the DTD index of
$1.0\la n \la 1.5$.
As a result of their frequent mergers, the amplitude of the GWB produced by the high-$z$ BBH population
can be $\Omega_{\rm gw}\simeq 1.48_{-1.27}^{+1.80}\times 10^{-9}$ at $f=25~{\rm Hz}$,
where the Advanced LIGO/Virgo detectors are the most sensitive.
The GWB spectrum is significantly flattened at $f\ga 20-30~{\rm Hz}$ from the value 
of $2/3$ and peaks inside the frequency window of the LIGO/Virgo observations.
Note that the flattened spectrum was predicted by previous studies \citep{Inayoshi_2016, Perigois_2020}
but the conclusion still holds even with the BBH properties updated from the LIGO-Virgo O3a observing run
and with the new Planck estimated value of $\tau_{\rm e}$.
This strong and characteristic GWB signal is detectable at the Advanced LIGO-Virgo design sensitivity.
The detection of this level of GWB would also indicate a major contribution of 
the high-$z$ BBH population to the local GW events.

We also consider a high-$z$ BBH population that follows a mass function more top-heavy than in the local universe,
motivated by the expected nature of high-$z$ star formation \citep[e.g.,][]{Hirano_2014},
BBH formation channels \citep{K14, Inayoshi_2017}, and possible subsequent growth processes 
via gas accretion \citep{Inayoshi_HE_2016,Tagawa_2020,Safarzadeh_2020}.
With a mass spectrum with a higher minimum mass, the peak frequency of the GWB moves to a lower value 
and thus the flattening of the spectrum becomes more prominent; namely, the spectral index becomes negative
at $f>20$ Hz. 
Even with the constraint from cosmic reionization, the GWB strength becomes as strong as 
$\Omega_{\rm gw}\simeq 4\times 10^{-9}$ at $f=20-25$ Hz.
A detection of such a unique spectrum with the design sensitivity would allow us to extract information on 
a top-heavy-like mass function of the high-$z$ BBH merger population.

Finally, we discuss the relation of the GWB amplitude and the local BBH merger rate in Fig.~\ref{fig:RBBH_GWB}.
In our fiducial case, where the merging-BBH formation efficiency is set to $f_{\rm BBH} \simeq 0.02$,
%$f_{\rm BBH}(\equiv \dot{\rho}_{\rm BBH}/\dot{\rho}_{\star}) \simeq 0.02$, 
the GWB produced by the high-$z$ BBHs is detectable at the design sensitivity and then those BBHs merge 
within the Advanced LIGO-Virgo detection horizon at a rate comparable to the observed rate.
If the high-$z$ BBHs form at a low efficiency of $f_{\rm BBH}\la 0.01$, 
the high-$z$ BBH population neither produces a GWB detectable at the design sensitivity
nor explains the local merger rate.
In this case, the low-$z$ BBH population dominates the local event rate and a GWB owing to the low-$z$ BBH population
would be detected at the envisioned A$^+$ sensitivity.
In addition, the current upper limit of the stochastic GWB obtained from the LIGO-Virgo O3a observing run
\citep{LIGO_GWB_2021} gives a constraint of $f_{\rm BBH}\la 0.04$.

%\vspace{3mm}
\acknowledgments
%We greatly thank Kenta Hotokezaka for the constructive discussion and KAGRA collaboration for . 
KI acknowledges support from the National Science Foundation of China 
(11721303, 11991052, 11950410493) and the National Key R\&D Program of 
China (2016YFA0400702).
KK acknowledges support from the JSPS KAKENHI Grant Numbers JP20K04010 and JP20H01904.
EV acknowledges support from NSF grant AST-2009309.
ZH acknowledges support from NASA through grant 80NSSC18K1093 and from
the National Science Foundation through grants 1715661 and 2006176.

\vspace{5mm}

\bibliography{ref}
\bibliographystyle{aasjournal}

\end{document}